\documentclass[numberedappendix,appendixfloats,twocolappendix]{emulateapj}

\usepackage{graphicx,amsmath,amsfonts,amssymb,subfigure,hyperref}
\newcommand\numberthis{\addtocounter{equation}{1}\tag{\theequation}}
\usepackage[titletoc,title]{appendix}

\begin{document}

\title{Reconciling the observed star-forming sequence with the observed stellar mass function}

\author{Joel Leja\altaffilmark{1}, Pieter G. van Dokkum\altaffilmark{1}, Marijn Franx\altaffilmark{2}, Katherine E. Whitaker\altaffilmark{3,4}}

\altaffiltext{1}{Department of Astronomy, Yale University, New Haven, CT 06511, USA} 
\altaffiltext{2}{Leiden Observatory, Leiden University, P.O. Box 9513, NL-2300 AA Leiden, The Netherlands}
\altaffiltext{3}{Astrophysics Science Division, Goddard Space Flight Center, Code 665, Greenbelt, MD 20771, USA}
\altaffiltext{4}{NASA Postdoctoral Program Fellow}
\slugcomment{Accepted for publication in ApJ}
\begin{abstract}
We examine the connection between the observed star-forming sequence (SFR $\propto$ $M^{\alpha}$) and the observed evolution of the stellar mass function between $0.2 < z < 2.5$. We find the star-forming sequence cannot have a slope $\alpha$ $\lesssim$ 0.9 at all masses and redshifts, as this would result in a much higher number density at $10 < \log(\mathrm{M/M_{\odot}}) < 11$ by $z=1$ than is observed. We show that a transition in the slope of the star-forming sequence, such that $\alpha=1$ at $\log(\mathrm{M/M_{\odot}})<10.5$ and $\alpha=0.7-0.13z$ ({Whitaker} {et~al.} 2012) at $\log(\mathrm{M/M_{\odot}})>10.5$, greatly improves agreement with the evolution of the stellar mass function. We then derive a star-forming sequence which reproduces the evolution of the mass function by design. This star-forming sequence is also well-described by a broken-power law, with a shallow slope at high masses and a steep slope at low masses. At $z=2$, it is offset by $\sim$0.3 dex from the observed star-forming sequence, consistent with the mild disagreement between the cosmic SFR and recent observations of the growth of the stellar mass density. It is unclear whether this problem stems from errors in stellar mass estimates, errors in SFRs, or other effects. We show that a mass-dependent slope is also seen in other self-consistent models of galaxy evolution, including semi-analytical, hydrodynamical, and abundance-matching models. As part of the analysis, we demonstrate that neither mergers nor hidden low-mass quiescent galaxies are likely to reconcile the evolution of the mass function and the star-forming sequence. These results are supported by observations from {Whitaker} {et~al.} (2014).
\end{abstract}
\keywords{
galaxies: evolution
}

\section{Introduction}
\begin{figure*}[th!]
\begin{center}
\includegraphics[bb=120 120 470 700,scale=0.74,angle=90]{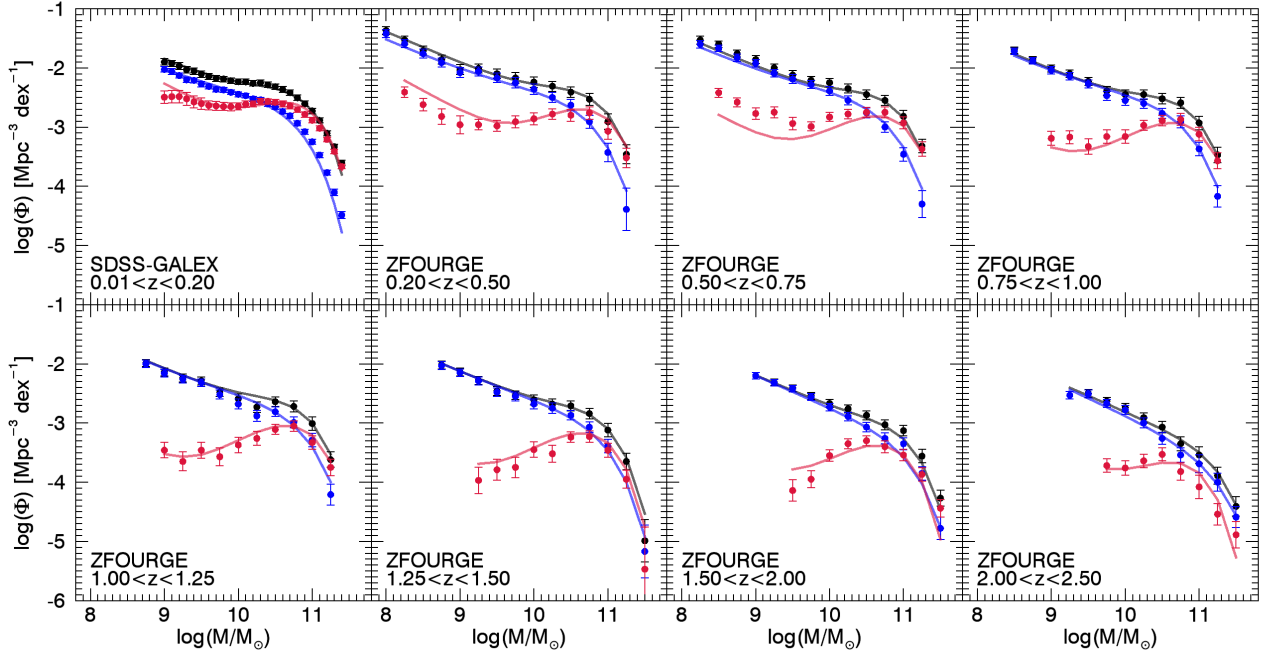}
\caption{The redshift evolution of the stellar mass function for all (black), star-forming (blue), and quiescent (red) galaxy populations. The filled circles are data from the ZFOURGE mass functions ({Tomczak} {et~al.} 2014) and SDSS-{\it GALEX} ({Moustakas} {et~al.} 2013), while the lines are the smooth model for $\phi(M,z)$, constructed as described in Section 2. The smooth model broadly reproduces the evolution of the stellar mass function between $0.2<z<2.5$.}
\label{schechparms}
\end{center}
\end{figure*}

The redshift evolution of the stellar mass function is a fundamental observable of galaxy evolution, as it directly measures the stellar mass buildup of galaxies. Recently, it has been measured with unprecedented precision by a number of wide, deep near-infrared surveys ({Muzzin} {et~al.} 2013; {Ilbert} {et~al.} 2013; {Moustakas} {et~al.} 2013; {Davidzon} {et~al.} 2013; {Tomczak} {et~al.} 2014). These surveys are revealing new frontiers in galaxy evolution; for example, the FourStar Galaxy Evolution Survey (ZFOURGE) is mass-complete to $\log(\mathrm{M/M_{\odot}})=9.0$ at $z=2$, for both star-forming and quiescent galaxies ({Tomczak} {et~al.} 2014). This impressive depth has revealed that simple Schechter fits are not a good representation of the mass function in the high-redshift Universe.

The observed relationship between the star formation rate (SFR) and stellar mass of star-forming galaxies (the ``star-forming sequence"), measuring the derivative of the mass buildup of galaxies, has also been of great interest in the literature ({Brinchmann} {et~al.} 2004; {Noeske} {et~al.} 2007; {Daddi} {et~al.} 2007; {Peng} {et~al.} 2010; {Karim} {et~al.} 2011; {Whitaker} {et~al.} 2012; {Guo}, {Zheng}, \& {Fu} 2013a; {Speagle} {et~al.} 2014). Recently, this has been measured robustly out to high redshift and over a wide variety of SFR indicators ({Oliver} {et~al.} 2010; {Wuyts} {et~al.} 2011; {Karim} {et~al.} 2011; {Whitaker} {et~al.} 2012), and further extended down to low stellar masses and star formation rates in the local Universe ({Huang} {et~al.} 2012). Much work has been done to bring these different SFR indicators into agreement ({Wuyts} {et~al.} 2011), and there has been some success in putting these many studies into a consistent framework ({Speagle} {et~al.} 2014).

While there has been great progress in reducing errors due to limited depth and field-to-field variations, the interpretation of these data is still subject to systematic uncertainties. Stellar mass measurements, particularly at higher redshifts, are uncertain by at least a factor of two, due to unknowns such as emission line contributions, star formation histories, dust content, and metallicities ({Marchesini} {et~al.} 2009; {Conroy}, {Gunn}, \& {White} 2009; {Behroozi}, {Conroy}, \&  {Wechsler} 2010; {Mitchell} {et~al.} 2013). While in principle these stellar mass measurements can be calibrated with dynamical masses, only a handful of galaxies at $z\sim2$ have reliable stellar velocity dispersions so far ({van de Sande} {et~al.} 2013; {Belli} {et~al.} 2014). Star formation rates also suffer from uncertainties, particularly in low-mass galaxies at high redshift, due to flux limits, systematic differences between SFR indicators, and selection effects ({Mitchell} {et~al.} 2014; {Speagle} {et~al.} 2014).

Given these systematic uncertainties, it is important to test whether the observed star-forming sequence and the observed evolution of the mass function are consistent with one another. A version of this phenomenological test was applied by {Bell} {et~al.} (2007), and has been implicitly performed by several studies since ({Drory} \& {Alvarez} 2008; {Peng} {et~al.} 2010; {Behroozi}, {Wechsler}, \&  {Conroy} 2013c). Most recently, {Weinmann} {et~al.} (2012) has showed that in order for the star-forming sequence to be consistent with the number density evolution of low-mass galaxies at $z<1$, either (1) for SFR $\propto$ M$^\alpha$, $\alpha$ must be greater than 0.9, or (2) the rate of destruction by mergers must be substantial. This analysis was limited to low redshift, however, thus missing the peak of the cosmic star formation rate density at $z$$\sim$2. Extention of this analysis to higher redshift requires accurate measurements of the number density of low-mass quiescent galaxies, which have only recently been made possible by the ultra-deep ZFOURGE mass functions ({Tomczak} {et~al.} 2014).

We use the ZFOURGE mass functions to take a fresh look at the consistency between the stellar mass function and the star-forming sequence between $0.2<z<2.5$. We compare this to the star-forming sequence from {Whitaker} {et~al.} (2012), which has been mapped with deep medium-band NIR imaging and consistent UV+IR SFR indicators between $0<z<2.5$.

The layout of the paper is as follows. In Section 2, we construct a smooth analytical description of the redshift evolution of the stellar mass function. In Section 3, we build a model to compare the growth of the mass function implied by different low-mass extrapolations of the star-forming sequence to the observed growth of the mass function, and based on this comparison, postulate a new functional form for the star-forming sequence. Section 4 discusses the implications of our results and the remaining uncertainties, and the conclusion is in Section 5.

We use a standard $\Lambda$CDM cosmology, with $\Omega_M$ = 0.3, $\Omega_{\Lambda}$ = 0.7, and $h=0.7$, and use a Chabrier initial mass function (IMF).

\section{The observed evolution of the mass function}
\label{massmod}
We adopt mass functions from the ZFOURGE survey, measured between $0.2 < z < 2.5$  ({Tomczak} {et~al.} 2014). ZFOURGE is the deepest measurement of the stellar mass function to date, and makes use of ground-based near-infrared medium-bandwidth filters which improve the accuracy of photometric redshifts ({van Dokkum} {et~al.} 2009). The survey also incorporates {\it HST} imaging from the Cosmic Assembly Near-infrared Deep Extragalactic Legacy Survey (CANDELS) ({Grogin} {et~al.} 2011; {Koekemoer} {et~al.} 2011). ZFOURGE use the CANDELS $H_{160}$ filter as their detection criteria. ZFOURGE imaging covers an area of 316 arcmin$^2$, and the ZFOURGE catalogs include data from the wider but shallower NEWFIRM Medium-Band Survey (NMBS; {Whitaker} {et~al.} 2011) to tighten constraints at the massive end. Star-forming and quiescent galaxies are separated by their rest-frame UVJ colors as described in {Williams} {et~al.} (2009).

We supplement the ZFOURGE mass functions with low-redshift mass functions measured from SDSS-$GALEX$ data ({Moustakas} {et~al.} 2013). These cover the redshift range $0.02<z<0.2$, and have $ugriz$ photometry and spectroscopic redshifts from the SDSS Data Release 7 ({Abazajian} {et~al.} 2009). They include $JHK_s$ photometry from the 2MASS Extended Source Catalog ({Jarrett} {et~al.} 2000) and photometry at 3.4 and 4.6 $\mu$m from the WISE All-Sky Data Release ({Wright} {et~al.} 2010). Star-forming and quiescent galaxies are separated via UV luminosity as measured by the {\it Galaxy Evolution Explorer} ($GALEX$; {Martin} {et~al.} 2005).
\begin{figure*}[th!]
\begin{center}
\includegraphics[bb=160 50 410 700,scale=0.65,angle=90]{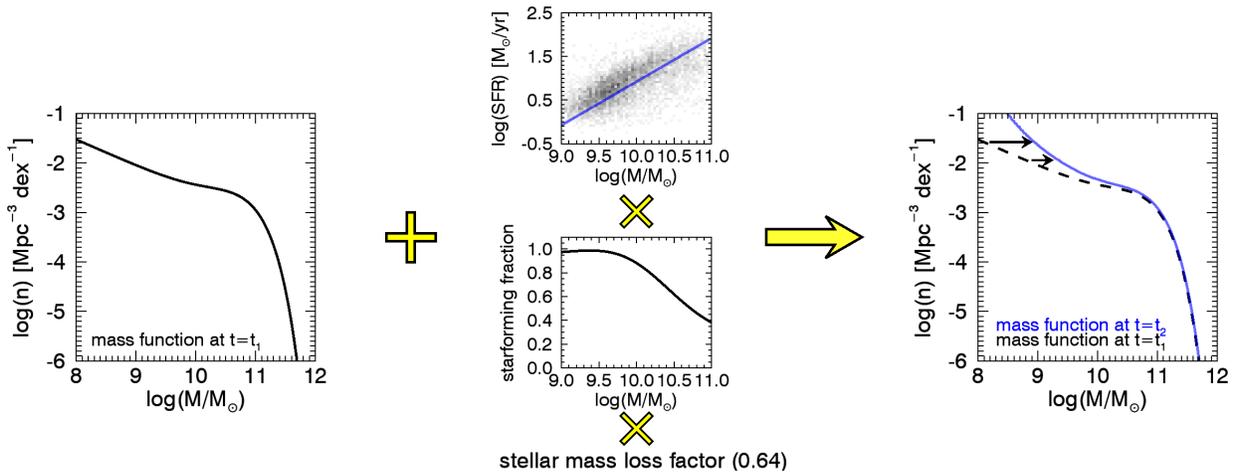}
\caption{A schematic representation of the model adopted to predict the evolution of the stellar mass function due to the star-forming sequence. We calculate the mass added to each bin of the mass function between time steps by multiplying the SFR-stellar mass relationship by the fraction of star-forming galaxies as a function of stellar mass, and by a constant factor of 0.64 to correct for passive stellar evolution. The addition of mass via star formation will shift the stellar mass function ``to the right", whereas mergers will shift it ``down and to the right".}
\label{schematic}
\end{center}
\end{figure*}

We use these data to construct an analytical description of the evolution of the stellar mass function with time. We aim to smoothly parameterize the redshift evolution of the stellar mass function as $\phi(M,z)$. Historically, the stellar mass function has been fit with a Schechter function ({Schechter} 1976). Recently, however, deeper measurements of the stellar mass function have shown that a double Schechter function is necessary to describe the steepening of the slope at masses below $10^{10}$ M$_{\odot}$ ({Baldry}, {Glazebrook}, \&  {Driver} 2008; {Li} \& {White} 2009; {Drory} {et~al.} 2009; {Moustakas} {et~al.} 2013; {Muzzin} {et~al.} 2013; {Ilbert} {et~al.} 2013; {Tomczak} {et~al.} 2014). The logarithmic form of the double Schechter function is:
\begin{align*}
\label{dblschech}
\phi(M)dM =& \mathrm{ln}(10)\exp{(-10^{(M-M^*)})}10^{(M-M^*)} \\
                   &[\phi_1^* 10^{(M-M^*)\alpha_1}+\phi_2^* 10^{(M-M^*)\alpha_2}]dM \numberthis
\end{align*}
where, in this equation, M = log(M/M$_{\odot}$) and M$^*$ = log(M$^*$/M$_{\odot}$). While in principle each component of the double Schechter could have a separate M$^*$, the data are consistent with a single value of M$^*$ for each component of the double Schechter function.

It is difficult to simply use the best-fit Schechter parameters to construct a smooth model for the evolution of the mass function, however, due to degeneracy between parameters in a double Schechter function: an unconstrained chi-squared minimization leads to Schechter parameters which do not evolve smoothly with redshift. We have found that simply interpolating in redshift between the best-fit Schechter parameters will introduce spurious increases and decreases in number density between observed redshift windows.

To avoid these numerical artifacts, we re-fit constrained double Schechter functions separately to the observed star-forming, quiescent, and total mass functions in each redshift window. We limit degenerate solutions by fixing the faint-end slopes to the best-fit values at $0.5<z<0.75$ from {Tomczak} {et~al.} (2014), which are: 
\begin{align*}
\mathrm{{\bf total}}: \alpha_{1} &=  -0.39 \\
\alpha_{2} &=  -1.53 \\
\mathrm{{\bf quiescent}}:\alpha_{1} &=  -0.10 \\
\alpha_{2} &=  -1.69 \\
\mathrm{{\bf star-forming}}:\alpha_{1} &=  -0.97 \\
\alpha_{2} &=  -1.58
\end{align*}
We follow {Drory} \& {Alvarez} (2008) in fitting second-order polynomials to the redshift evolution of the remaining best-fit Schechter parameters, namely $\phi_1$, $\phi_2$, and $M^*$. We then re-fit $\phi_1,\phi_2,$ and $M^*$, constraining them to be within 40\% of the best-fit second-order polynomials. We perform this fit iteratively, constraining the best-fit $\phi_1$, $\phi_2$, and $M^*$ to be increasingly close to the second-order polynomials during each fit. 

The goals of this procedure are to: (1) reproduce the evolution of the observed mass function in the observed redshift windows, and (2) to enforce smooth, monotonic evolution of the mass function between observed redshift windows. This iterative approach maximizes the effectiveness of the polynomial fits in building a smooth model for redshift evolution of the stellar mass function, though we recognize that it does not necessarily guarantee a unique solution.

The resulting redshift evolution of the Schechter parameters is:
\begin{align*}
\mathrm{{\bf total}}: \log{(\phi_1)} &= -2.46+0.07z-0.28z^2 \\
\log{(\phi_2)} &= -3.11-0.18z-0.03z^2 \\
\log{(M^*/M_{\odot})} &=10.72-0.13z+0.11z^2 \\
\mathrm{{\bf quiescent}}: \log{(\phi_1)} &= -2.51-0.33z-0.07z^2\\
\log{(\phi_2)} &= -3.54-2.31z+0.73z^2\\
\log{(M^*/M_{\odot})} &\equiv 10.70\\
\mathrm{{\bf star-forming}}: \log{(\phi_1)} &=-2.88+0.11z-0.31z^2\\
\log{(\phi_2)} &= -3.48+0.07z-0.11z^2\\
\log{(M^*/M_{\odot})} &=10.67-0.02z+0.10z^2
\end{align*}
As we find no significant redshift evolution in the quiescent $M^*$, we fix it to its average value. 

The parameterized mass function growth, $\phi(M,z)$, is compared directly with the measured ZFOURGE mass functions in Figure \ref{schechparms}. There is overall good agreement with the observed mass function. Noticeably, the observed number density of low-mass quiescent galaxies is underpredicted by $\sim0.3$ dex at $0.5<z<0.75$ and overpredicted by $\sim 0.2$ dex at $0.2<z<0.5$. The data show negligible evolution in the number density of low-mass quiescent galaxies between these two redshift windows. However, there is still significant evolution from $z=1$ to $z=0.35$, and from $z=0.625$ to $z=0.1$, and the smooth evolution in the model reflects the broader trend of declining number density in low-mass quiescent galaxies with increasing redshift. In practice, this offset has a negligible effect on our conclusions: the quiescent mass function is only used to calculate the fraction of star-forming galaxies as a function of mass, and star-forming galaxies dominate at low masses regardless of the normalization offset. 

In the following analysis, we extrapolate the stellar mass function to below the nominal stellar mass completeness when necessary. The assumed shape of the stellar mass function below the observed completeness limit does not have a significant effect on our results, however.
\begin{figure*}
\begin{center}
\includegraphics[bb=90 100 470 725,scale=0.7,angle=90]{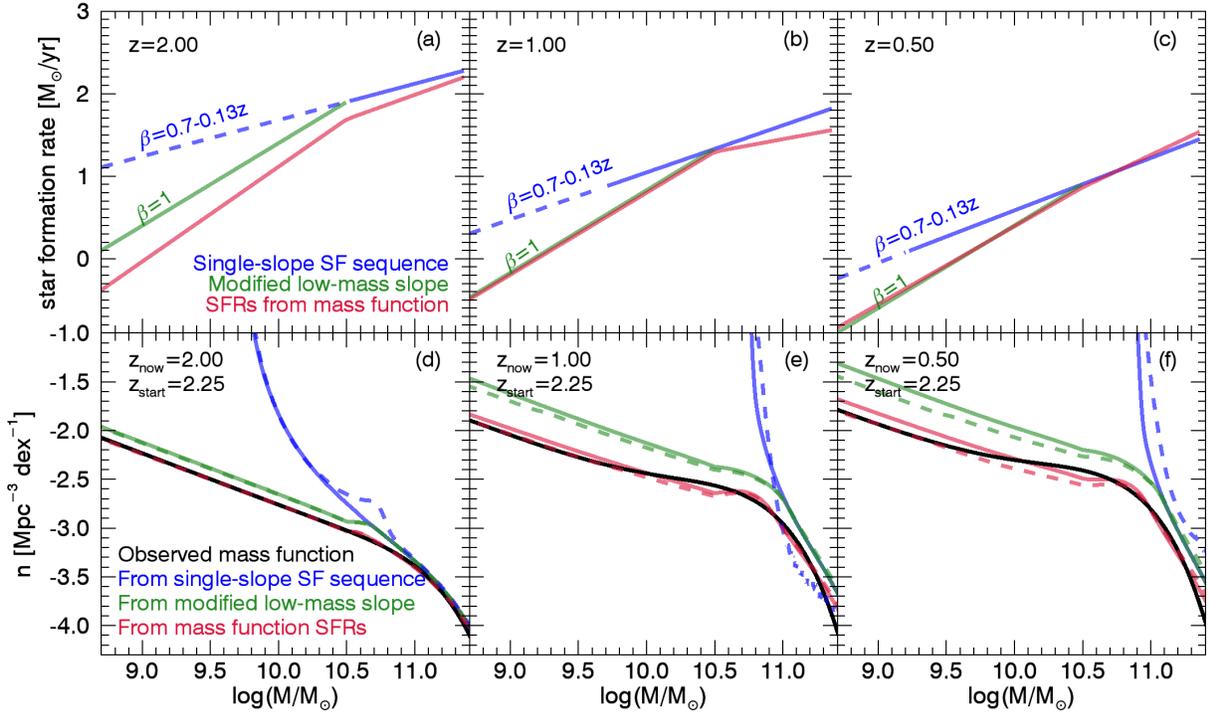}
\caption{Panels a-c show different adopted relationships between stellar mass and star formation rate. The observed star-forming sequence (blue) is from {Whitaker} {et~al.} (2012), with the extrapolation below their mass-completeness limit marked with a dashed line. The modified low-mass slope (green) uses the {Whitaker} {et~al.} (2012) slope above log(M/M$_{\odot}$) = 10.5, and a slope of unity below log(M/M$_{\odot}$) = 10.5. The SFR from the ZFOURGE mass functions is constructed as described in Section \ref{zfourgesfr} for stellar masses above the ZFOURGE mass completeness limits, and is fit with a broken power law. Panels d-f show the observed ZFOURGE mass function (black), along with the mass functions grown by the corresponding star formation rates from $z=2.25$. The growth of the mass function with mergers applied is shown with a dashed line. Without a steep low-mass slope for the star-forming sequence, the number density of galaxies is quickly overpredicted.}
\label{mainplot}
\end{center}
\end{figure*}
\section{The evolution of the mass function as implied by the star forming sequence}
We now compare the evolution of the mass function, as implied by the star-forming sequence, to the observed evolution of the mass function, as parameterized in Section \ref{massmod}. We begin by describing the model which connects the star-forming sequence to the growth of the mass function. We also include a simple model for the effect of galaxy mergers on the evolution of the mass function. We then examine the effects of applying different star-forming sequences to the mass function.
\subsection{Model for the growth of the mass function}
\label{mgrowthmod}
We implement a model to describe the time evolution of the mass function due to the observed star-forming sequence. A schematic of this model is shown in Figure \ref{schematic}. We also include a simple model for galaxy mergers.
\subsubsection{Growth of the mass function due to star formation}
At each redshift step, mass is added to the mass function by star formation. Star formation will cause the mass function to shift to the ``right" with time. At fixed mass, galaxies have a range of star formation rates ({van Dokkum} {et~al.} 2011). However, when describing the evolution of the mass function, the complex distribution of star formation rates can be reduced to the average star formation rate as a function of mass (discussed further in Section \ref{scatter}). 

In general, each bin of the mass function contains a mix of star-forming and quiescent galaxies. Adopting this division, the rate of mass addition due to star formation in each bin of the mass function is given by:
\begin{equation}
\dot{M}_{avg} = \frac{\phi_{sf}\dot{M}_{sf} + \phi_{qu}\dot{M}_{qu}}{\phi_{sf}+\phi_{qu}}
\end{equation}
with $\dot{M}_{sf}$ the average star formation rate of star-forming galaxies and $\dot{M}_{qu}$ the average star formation rate of quiescent galaxies. We assume that quiescent galaxies have a negligible rate of star formation and set $\dot{M}_{qu}$ to zero. This assumption is justified by observations indicating that UVJ-selected quiescent galaxies have average star formation rates that are at least 20-40 times lower than UVJ-selected star-forming galaxies, and thus can be safely neglected ({Fumagalli} {et~al.} 2013).

We define $f_{sf}$, the fraction of star-forming galaxies, to be the following:
\begin{equation}
f_{sf}(M,z) = \frac{\phi(M,z)_{sf}}{\phi(M,z)_{qu} + \phi(M,z)_{sf}}
\end{equation}
We calculate $f_{sf}(M,z)$ directly from our smoothed model for the evolution of the mass function, and fit it with:
\begin{equation}
\label{f_sf}
f_{sf} = f_0 - (f_0-0.2)\tanh[a(\log(\mathrm{M/M_{\odot}})-b)]
\end{equation}
The fit is restricted to $9<\log(\mathrm{M/M_{\odot}})<11$, where the mass function is mass-complete and the effects of cosmic variance are minimized at the high-mass end ({Tomczak} {et~al.} 2014). By construction, the adopted star-forming fraction asymptotes at high masses to $f_{sf} = 0.2$, consistent with studies of BCGs in the local Universe ({Bauer} {et~al.} 2013; {Oliva-Altamirano} {et~al.} 2014). At low masses, the star-forming fraction asymptotes to $f_0$; the typical best-fit value of $f_0$ is $\sim0.9$.

We take $\dot{M}_{sf}$ to be:
\begin{equation}
\dot{M}_{sf}(M,z) = SFR(M,z)\times(1-R)
\end{equation}
$SFR(M,z)$ is the star formation rate implied by the star-forming sequence. $R$ is the fraction of mass ejected from a stellar population during the course of passive stellar evolution, primarily due to winds and outflows. In general, $R=R(t)$, starting at $R=0$ when the stellar population is formed and ending at $R=0.36$ after $\sim10$ Gyr (assuming a Chabrier IMF). However, as most of the mass loss occurs within the first hundred million years, we approximate mass loss as instantaneous, fixing $R$ to $0.36$ (see Section \ref{mloss} for further discussion of this approximation). The final equation for the average growth rate of the mass function due to star formation is thus:
\begin{equation}
\label{finaleq}
\dot{M}_{avg} = SFR(M,z)\times f_{sf}(M,z) \times (1-R)
\end{equation}
\subsubsection{Growth of the mass function due to mergers}
The framework described so far models the growth of the mass function solely due to star formation. The other physical process that can change the number density of galaxies is galaxy mergers. Specifically, mergers affect the mass function in two ways: (1) they directly decrease the number density of galaxies, and (2) they contribute to the stellar mass growth of galaxies. 

We model both of these effects. We measure the merger rate measured directly from the {Guo} {et~al.} (2013b) semi-analytical model (SAM) based on the Millenium-II simulation ({Boylan-Kolchin} {et~al.} 2009). At each snapshot in redshift, we measure the rate at which galaxies merge with more massive galaxies than themselves, which is the rate of ``destruction" by mergers. This is shown as a function of mass in Figure \ref{samdirect} in Appendix A. For low-mass galaxies, this rate varies by $1-2\%$ per snapshot in redshift, but broadly declines from $\sim$7\% per Gyr at $z\sim2$ to $\sim$3\% per Gyr in the local Universe. This is consistent with observed destruction rates of $\sim$10\% per Gyr ({Bridge}, {Carlberg}, \&  {Sullivan} 2010; {Lotz} {et~al.} 2011), though other studies which determine merger rates based on galaxy morphology find destruction rates that are 3-5 times higher ({Conselice} 2014). We show in Section \ref{otherexp} that our conclusions are robust to changing the destruction rate.

We interpolate in redshift and stellar mass to apply this measured destruction rate directly to the evolution of the mass function. In order to enforce stellar mass conservation, we assume that all mergers have a mass ratio of 1:10, and correspondingly increase the stellar mass of more massive galaxies. As we show below, including mergers changes the evolution of the mass function at low masses by only $0.1-0.2$ dex in number density from $z=2.25$ to $z=0.5$. In Appendix A, we show that instead using the measured mass ratios from the {Guo} {et~al.} (2013b) SAM has little effect on our results. We note that using measured mass ratios from the {Guo} {et~al.} (2013b) SAM no longer enforces mass conservation in our model, as the stellar mass functions within the semi-analytical model do not match those from observations.

\subsection{Modeling the growth of the mass function with different star-forming sequences}
\label{dif_sfseq}
In this section, we compare the effects of using three different star-forming sequences to grow the mass function. We start the simulation at $z=2.25$ with the observed mass function. In each time step, we then evolve the mass function according to the model described in Section \ref{mgrowthmod}, and compare it to the observed mass function. The resulting growth of the mass function, with and without mergers, is illustrated in Figure \ref{mainplot}, along with the corresponding star-forming sequences. We describe the adopted star-forming sequences and their corresponding effect on the evolution of the mass function below.

\begin{figure}[th!]
\begin{center}
\includegraphics[bb=200 100 415 650,scale=0.4]{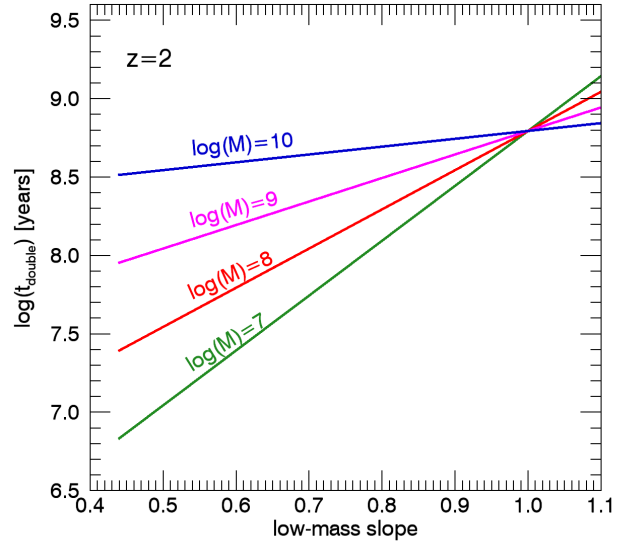}
\caption{The timescale over which a galaxy will double its stellar mass due to star formation is plotted as a function of the low-mass slope of the star-forming sequence from Equation \ref{bpl_eqn}. Mass loss is taken to be a constant factor $R=0.36$ for consistency with the rest of the study. For flat low-mass slopes, low-mass galaxies grow at an extremely rapid rate at $z=2$, inconsistent with the observed growth of the mass function.}
\label{tdouble}
\end{center}
\end{figure}

\begin{figure*}[th!]
\begin{center}
\includegraphics[bb=190 50 380 700,scale=0.65,angle=90]{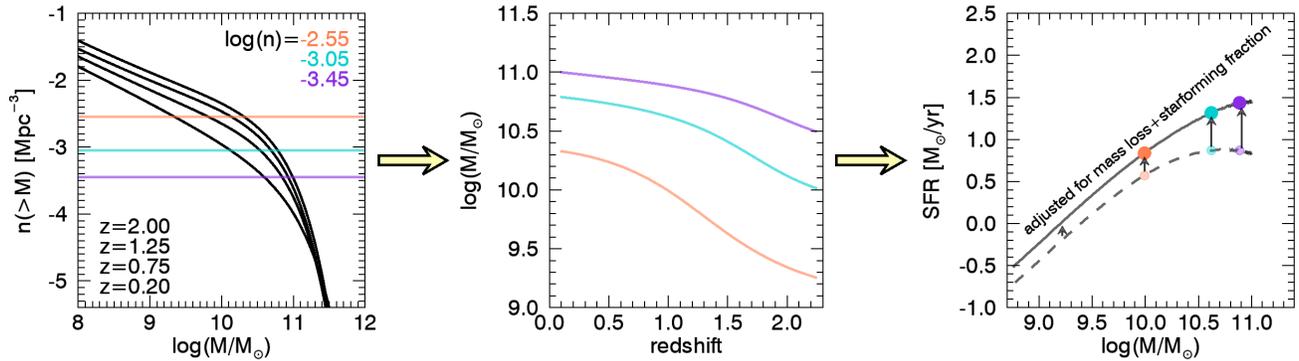}
\caption{This figure describes the procedure to construct a star-forming sequence which is precisely consistent with the growth of the stellar mass function, under the assumptions that mergers are negligible and there is no scatter in star formation rates. In this case, galaxies evolve along lines of constant number density ({Leja}, {van Dokkum}, \& {Franx} 2013). For a given number density, this allows us to derive relationship between redshift and stellar mass from $\phi(M,z)$. It is straightforward to convert this into $\dot{M}(M)$, as shown schematically above. We convert $\dot{M}(M)$ into star formation rates by applying the inverse of the corrections for quiescent galaxies and passive stellar evolution shown in Figure \ref{schematic}.}
\label{schematic2}
\end{center}
\end{figure*}

\subsubsection{The observed star-forming sequence}
\label{obs_sfseq}
We first use the observed star-forming sequence from {Whitaker} {et~al.} (2012) to model the growth of the stellar mass function. {Whitaker} {et~al.} (2012) measure stellar masses and photometric redshifts from NMBS, which combines photometry in 5 near-infrared medium-band filters with publicly-available imaging from 0.15-8$\mu$m ({Whitaker} {et~al.} 2011). They use a combination of $Spitzer$-MIPS and UV imaging to determine SFRs in a uniform manner between $0<z<2.5$. Star-forming galaxies are defined by their rest-frame UVJ colors. We note that the ZFOURGE stellar mass function uses the same color-color cut and also uses the NMBS imaging in derivation of their stellar masses. Thus, the mass function and the star formation rates are measured, to a large extent, from the same galaxies, which will minimize the effect of systematic differences in stellar masses between these two data sets.

{Whitaker} {et~al.} (2012) provide the following single power law fit to the median\footnote{We note that our model requires the {\it average} SFR of star-forming galaxies; however, the observed average SFR is sensitive to outliers and AGN contamination, so here we assume that the median is a good approximation of the true average. In practice, direct examination of the NMBS data reveals the distribution of star formation rates in star-forming galaxies is such that the average is $\sim0.1$ dex higher than the median. The observed star formation rates used in our model may be slightly too low, if the measured average is closer to the true average than the measured median.} SFR of star-forming galaxies:
\begin{equation}
\mathrm{log(SFR)} = \alpha(z) (\log(\mathrm{M/M_{\odot}})-10.5)+\beta(z),
\end{equation}
with $\alpha(z) = 0.70-0.13z$ and $\beta(z) = 0.38+1.14z-0.19z^2$. Due to incompleteness at low masses and low star formation rates, the slope of the star-forming sequence at low stellar mass is not well-constrained in this study. For reference, the mass-complete limit for NMBS is $\log(\mathrm{M/M_{\odot}}) \sim10.65$ at $z=2.25$ and $\log(\mathrm{M/M_{\odot}}) \sim10$ at $z=1.25$, and the SFR-complete limit for {Whitaker} {et~al.} (2012) is $\sim40$ M$_{\odot}$/yr at $z=2.25$ and $\sim15$ M$_{\odot}$/yr at $z=1.25$. We first assume that the slope does not change as a function of mass. In the following subsection, we explore the effects of altering this assumption.

This extrapolation of the observed slope results in a dramatic rate of growth for low-mass galaxies, visible in Figure \ref{mainplot}. By $z=2$, only $\sim350$ Myr after the simulation has started, the number density of galaxies with $\log(\mathrm{M/M_{\odot}})$ = 9 is overpredicted by a factor of 100. This offset decreases as stellar mass increases: at $\log(\mathrm{M/M_{\odot}})$ = 10.5, the offset is only a factor of 2. The dramatic growth of these galaxies is a result of the relatively flat slope of the star-forming sequence. At $z=2$, the observed slope of the star-forming sequence is 0.44, meaning that the naive mass-doubling time scales as $t_{double} = M/SFR \propto M^{-0.56}$. With the observed normalization of the star-forming sequence, a galaxy at $\log(\mathrm{M/M_{\odot}})$= 8 has a doubling time of just 21 million years. The very rapid assembly of the mass function is thus a direct result of the implied high star formation rates at low masses.

When mergers are turned on, the rapid growth of low-mass galaxies causes a corresponding strong growth in high-mass galaxies. The merger growth rate is very rapid, such that by $z=1$, there is actually a deficit of high-mass galaxies, as they have all grown to $>10^{11.5}$ M$_{\odot}$. This smooths out by $z=0.5$, as the lower-mass galaxies grow due to star formation and mergers.

This implied rapid growth of low-mass galaxies is consistent with previous analyses from the literature ({Drory} \& {Alvarez} 2008; {Weinmann} {et~al.} 2012). {Weinmann} {et~al.} (2012) examines the difference in the evolution of low-mass galaxies between observations and hydrodynamical+semi-analytical models of galaxy evolution. In the process of this comparison, they model the evolution of the mass function between $0<z<0.9$ by varying the slope of the star-forming sequence. They conclude that the number density of galaxies with masses between 9.27 $< \log(\mathrm{M/M_{\odot}}) <$ 9.77 is consistent with the observed number density if the slope of the star-forming sequence is steeper than $\sim0.9$, though their model does not include corrections for quiescent galaxies. They suggest that either high merger rate or a steep star-forming sequence can reconcile the mass function with the star-forming sequence. {Drory} \& {Alvarez} (2008) use $I$-band selected mass functions and average UV star formation rates to estimate the galaxy merger rates, calculated by subtracting the evolution of the mass function due to star formation from the observed evolution of the mass function. They conclude that the net merger rate of low-mass galaxies must be high to be consistent with the growth of the mass function: 14\% per 100 Myr at $\log(\mathrm{M/M_{\odot}})=9$. They note that this result is sensitive to the slope of the star-forming sequence (they adopt a slope of 0.6) and to completeness estimates at low stellar mass.

The merger rate calculated from the {Guo} {et~al.} (2013b) SAM for galaxies with $\log(\mathrm{M/M_{\odot}}) \sim 8$ is only $\sim$ 0.5-0.7\% per 100 Myr, more than order of magnitude lower than the net merger rate estimated in {Drory} \& {Alvarez} (2008). Application of the SAM merger rate to the evolution of the mass function is indicated by the dashed lines in Figure \ref{mainplot}. This rate of galaxy mergers cannot compensate for the rapid growth of low-mass galaxies implied by the observed star-forming sequence.

\subsubsection{A modified low-mass slope}
\label{bpl_sec}
We next explore the effect of modifying the slope at low stellar masses. We postulate a simple broken power law:
\begin{equation}
\label{bpl_eqn}
 \log(\mathrm{SFR}) = \left\{ 
  \begin{array}{l l}
     \alpha(\log(\mathrm{M/M_{\odot}})-10.5)+\beta & M > 10^{10.5}M_{\odot}\\
     (\log(\mathrm{M/M_{\odot}})-10.5) + \beta & M \le 10^{10.5}M_{\odot}
  \end{array} \right.
\end{equation}
with the high-mass slope and normalization fixed to the {Whitaker} {et~al.} (2012) values. The low-mass slope is taken to be unity.

As can be seen in Figure \ref{mainplot}, adopting this star-forming sequence effectively removes the rapid growth of low-mass galaxies. This is due to a steeper low-mass slope, which in turns implies longer mass-doubling timescales ($t_{double}$) for low-mass galaxies. This is illustrated in Figure \ref{tdouble}, which shows the mass-doubling timescale at $z=2$ as a function of low-mass slope in Equation \ref{bpl_eqn}. Extrapolating the observed high-mass slope of 0.44 at $z=2$ to low masses results in a doubling time of less than 10 million years for a galaxy of mass $\log(\mathrm{M/M_{\odot}})$ = 7, compared to $\sim600$ Myr for a slope of unity.

Thus, adopting a slope of unity at low masses results in evolution of the mass function which compares much more favorably with the observed mass function. However, there remains an offset from the observed mass function of $0.2-0.4$ dex at all masses. This offset originates at $z=2$ and persists to lower redshifts.

\subsubsection{SFRs from the mass function}
\label{zfourgesfr}
We now generalize the result of Section \ref{bpl_sec} by constructing a star-forming sequence which exactly reproduces the observed evolution of the mass function. In our simple model with no scatter in star formation rates and a slow or nonexistent merger rate, galaxy populations will evolve along lines of constant number density ({Leja} {et~al.} 2013). We take a range of evenly-spaced slices in number density between $-4.3 < \log(n_{cum}/\mathrm{Mpc}^{3}) < -1.7$ and calculate $\dot{M}(M)$ directly from $\phi(M,z)$. We apply the inverse of the corrections for star-forming fraction and mass loss described above to turn this into a star-forming sequence (see Figure \ref{schematic2}). 

We then perform a least chi-square fit to the implied star formation rates with the following function:
\begin{equation}
\label{brokenpowerlaw}
 \log(\mathrm{SFR}) = \left\{ 
  \begin{array}{l l}
     \alpha_{1} (\log(\mathrm{M})-\log(\mathrm{M_t})) + \beta & \quad M > M_{t}\\
     \alpha_{2} (\log(\mathrm{M})-\log(\mathrm{M_t})) + \beta & \quad M \le M_{t}
  \end{array} \right.
\end{equation}
We restrict the fit to masses above the ZFOURGE stellar mass completeness limit, and below $\log(\mathrm{M/M_{\odot}})$ = 11, where star formation is expected to be the dominant mode of stellar mass growth ({Drory} \& {Alvarez} 2008; {Conroy} \& {Wechsler} 2009; {Leitner} 2012). 

This model has four parameters: the transition mass $M_t(z)$, the high-mass slope $\alpha_1(z)$, the low-mass slope $\alpha_2(z)$, and the normalization $\beta(z)$. As the model is not very sensitive to the location of the transition mass, we fix $M_t(z)=10.5$ in all fits. This lack of sensitivity comes from the fact that, in reality, the slope changes smoothly with mass in the transition region. However, lacking the theoretical basis for a more physically motivated functional form, we prefer a double power law for simplicity.

\begin{figure}
\begin{center}
\includegraphics[bb=80 75 550 690, scale=0.6]{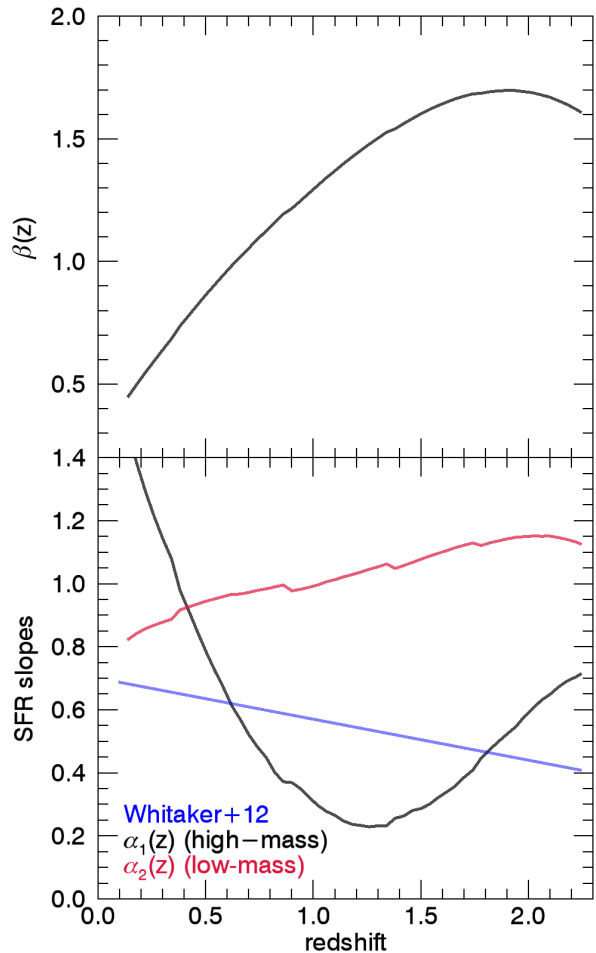}
\caption{The best-fit slopes and normalization for the broken power law fit to the star formation rates derived from the evolution of the ZFOURGE mass functions. In general, the growth of the mass function requires a star-forming sequence with a relatively steep slope at low masses, and a more shallow slope at high masses. At low redshift, the high-mass slope steepens substantially. This is not corroborated by observations and may be related to the effects of mergers at higher masses-- see the discussion at the end of Section \ref{zfourgesfr}.}
\label{parms}
\end{center}
\end{figure}

The best-fit redshift evolution of these parameters is shown in Figure \ref{parms}. In general, at higher redshifts, the mass function prefers a steep low-mass slope and a shallow high-mass slope. At low redshifts, the high-mass and low-mass slopes head towards convergence. This may be a sign that the star-forming sequence locally is well-fit by a single power law down to observed stellar mass completeness limits. The periodic structure visible in $\alpha_2$ is a result of the ZFOURGE mass completeness limits. 

We show in Figure \ref{mainplot} that this model correctly describes the growth of the observed mass function. There do exist some small discrepancies. At $z=0.5$, the number density of galaxies with masses between $8.8 < \log(\mathrm{M/M_{\odot}}) <  9.7$ is overpredicted by $\sim0.1$ dex. These low-mass galaxies were below the mass-complete limit at the start of the simulation, and so both their original number densities and their expected growth rates came from fits to galaxies at higher masses, extrapolated to low mass. Thus, their low-redshift number densities are not expected to match the observed number densities exactly. The other noticeable difference is around the transition mass of $\log(\mathrm{M/M_{\odot}})) = 10.5$, where the number density is under-predicted below and over-predicted above by $0.05-0.1$ dex. This is a natural result of fitting a bimodal double-power law to an intrinsically curved distribution of star formation rates: the star formation rates directly below the transition mass are slightly over-predicted, and the star formation rates directly above the transition mass are slightly under-predicted. A more physically motivated functional form will resolve these differences in the future.

Below $z\sim0.45$, the high-mass slope becomes considerably steeper than the low-mass slope. This is a direct result of the number density of high-mass galaxies increasing substantially at low redshift, an increase which is both intrinsic to the ZFOURGE mass functions and also arises naturally when ZFOURGE is combined with the SDSS-$GALEX$ mass functions\footnote{We note also that not all mass functions show a similar growth in the number density of galaxies at high stellar masses-- see {Moustakas} {et~al.} (2013).}. We caution that while our model interprets this as an increase in the star formation rate of star-forming massive galaxies at low redshift, this is not seen in observations (e.g., {Noeske} {et~al.} 2007). An alternate explanation is that mergers, both major and minor, are important in the evolution of massive galaxies and correspondingly in the evolution of the mass function at the high-mass end ({van Dokkum} 2005; {Bezanson} {et~al.} 2009; {Naab}, {Johansson}, \& {Ostriker} 2009; {van Dokkum} {et~al.} 2010; {Behroozi} {et~al.} 2013a; {Ferreras} {et~al.} 2014). Our simple prescription for mergers is insufficient to model the evolution of the high-mass end of the mass function in detail, particularly since while the net effect of mergers on the mass function may be small, the effect on individual galaxies may  be quite substantial ({Drory} \& {Alvarez} 2008).

We explicitly test this idea in Appendix A by deriving SFRs from the mass function {\it including} mass growth and destruction via mergers. The low-redshift increase in the high-mass slope then disappears. Instead, the high-mass slope drops rapidly at low redshift, becoming negative. This is because the growth via mergers is sufficient (or more than sufficient) to explain the growth at the high-mass end. We conclude that in regimes where the growth via star formation is sub-dominant (i.e., at low redshifts and high masses), the derived slope of the star-forming sequence from the stellar mass function is sensitive to the growth rate via mergers. This is not the case for the low-mass slope, but is the case for the high-mass slope. We thus suggest that observational studies are a more robust determinant of the high-mass slope than the analytical model, and do not find significant evidence that the high-mass slope as measured in, e.g., {Whitaker} {et~al.} (2012) is incorrect.

\begin{figure*}[t]
\begin{center}
\includegraphics[bb=0 80 570 700,scale=0.45]{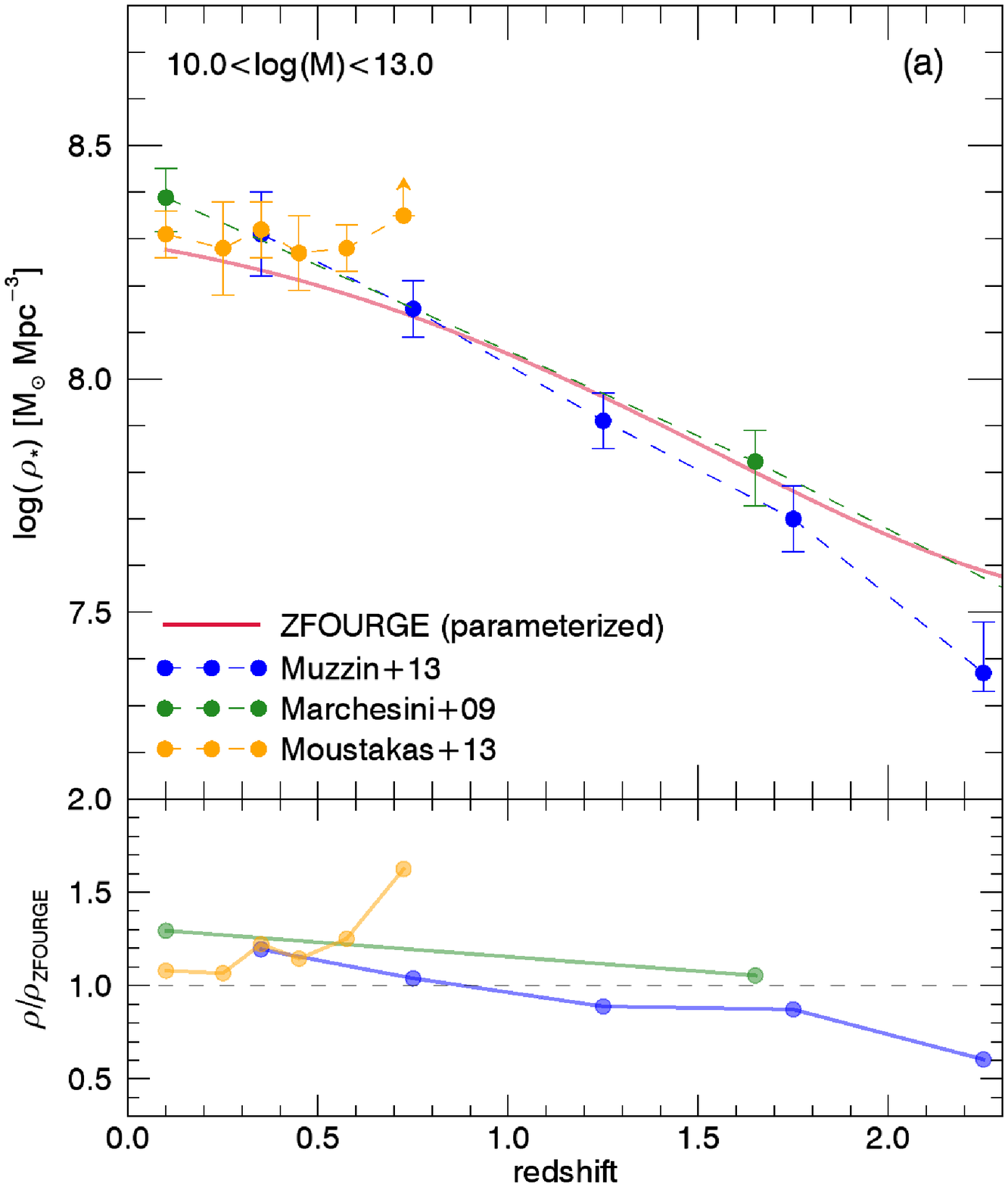}
\includegraphics[bb=30 80 570 700,scale=0.45]{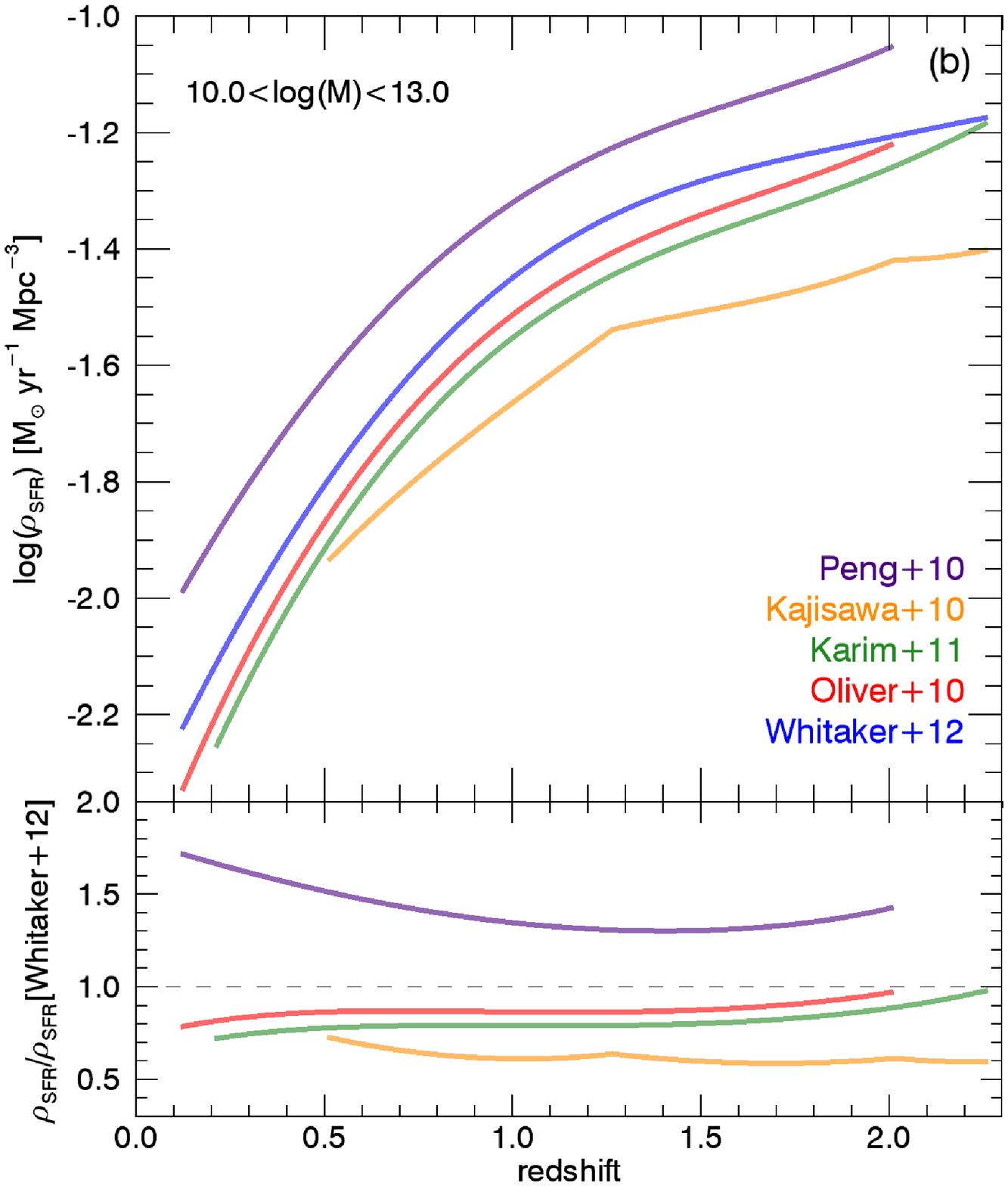}
\caption{Panel (a) shows the buildup of stellar mass density between $10 <$ $\log(\mathrm{M/M_{\odot}})$ $< 13$ as measured from multiple stellar mass functions. Below, the ratio of the measured stellar mass density relative to ZFOURGE is shown. The buildup of stellar mass varies between observations by a factor of $\sim2$, and ZFOURGE is comparable to other stellar mass functions in this respect-- though it does measure a higher mass density in low-mass galaxies. In panel (b), we show the global star formation rate in galaxies with $10 <$ $\log(\mathrm{M/M_{\odot}})$ $< 13$ as calculated from different observed star-forming sequences. Below, we show the ratio of the measured global star formation rate relative to that implied by the {Whitaker} {et~al.} (2012) star-forming sequence. The global star formation rates also vary by a factor of $\sim2$, and the {Whitaker} {et~al.} (2012) star-forming sequence is comparable to other observed star-forming sequences.}
\label{litcomp}
\end{center}
\end{figure*}
\section{Discussion}
\label{disc}

\subsection{A changing slope of the star-forming sequence}
In this study, we have demonstrated that a straightforward extrapolation of the observed star-forming sequence to low masses results in dramatic disagreement with the observed evolution of the mass function. The assumption of a single-slope starforming sequence implies mass-doubling times of 25 million years for a galaxy of mass $10^8$ M$_{\odot}$ at $z=2$, which is inconsistent with the observed growth of the mass function. We further demonstrate that this inconsistency is unlikely to be solved by either mergers or undiscovered populations of low-mass quiescent galaxies. Thus, we posit that the most likely explanation is a steeper slope for the star-forming sequence at low masses than is observed at high masses.

This is the first explicit test of the consistency between the star-forming sequence and the stellar mass function since {Bell} {et~al.} (2007), made possible in part by recent, accurate, and deep high-redshift measurements of the stellar mass function and the star-forming sequence ({Whitaker} {et~al.} 2012; {Tomczak} {et~al.} 2014). Another key advance in this study is demonstrating that neither mergers nor an unknown population of quiescent low-mass galaxies is likely to solve this problem. The results of this study are consistent with related joint analyses of the star-forming sequence and the stellar mass function ({Drory} \& {Alvarez} 2008; {Weinmann} {et~al.} 2012).

A key question is whether a steep low-mass slope, disjoint from a shallower, high-mass slope, is consistent with observations. The evidence presented in this study for a steep low-mass slope is most compelling at $z=2$ (see Figure \ref{mainplot}), due to the low mass-doubling timescales implied by a single-slope star-forming sequence (Figure \ref{tdouble}). Some studies have observed a downturn in the SFR-mass relationship in star-forming galaxies at high redshift, but this lies close to or below their completeness limits, preventing any strong conclusions ({Karim} {et~al.} 2011; {Whitaker} {et~al.} 2012). {Salim} {et~al.} (2007) finds a mass-dependent slope in their data before removing AGN, but the mass-dependent slope disappears after removing AGN from their data.

In a companion paper ({Whitaker} {et~al.} 2014), we present new observational evidence for a mass-dependent slope. {Whitaker} {et~al.} (2014) measures the slope of the star-forming sequence separately for galaxies with $\log(\mathrm{M/M_{\odot}})$ $>$ 10.2 and those with $\log(\mathrm{M/M_{\odot}})$ $>$ 10.2, finding that the low-mass slope is approximately unity, and remains stable within observational uncertainties between $0.5 < z < 2.5$, while the high-mass slope decreases from roughly 0.8 at $z=2.5$ to 0.2 at $z=0.5$. These findings are consistent across observational indicators for the star formation rate.

\subsection{Remaining discrepancies between star formation rates and stellar masses}
\label{discrep}
Even after adopting a steep slope for the star-forming sequence below $\mathrm{log(\mathrm{M/M_{\odot}})}=10.5$, there remains a $\sim$0.3 dex offset at $z=2$ between the growth of the mass function and the normalization of the star-forming sequence. This offset is present at all stellar masses, and largely disappears by $z=1$. This discrepancy may arise from problems in stellar mass estimates, star formation rate estimates-- or, most likely, both. Since the {Whitaker} {et~al.} (2012) starforming sequence and the ZFOURGE mass functions both use NMBS imaging to derive stellar masses, the difference is very unlikely to stem from systematic differences between these two studies. This cannot be fixed by adjusting the merger rates, as the high-mass end already overproduces stars even when neglecting growth via mergers.

\begin{figure*}[t]
\begin{center}
\includegraphics[bb=180 120 410 700,scale=0.74,angle=90]{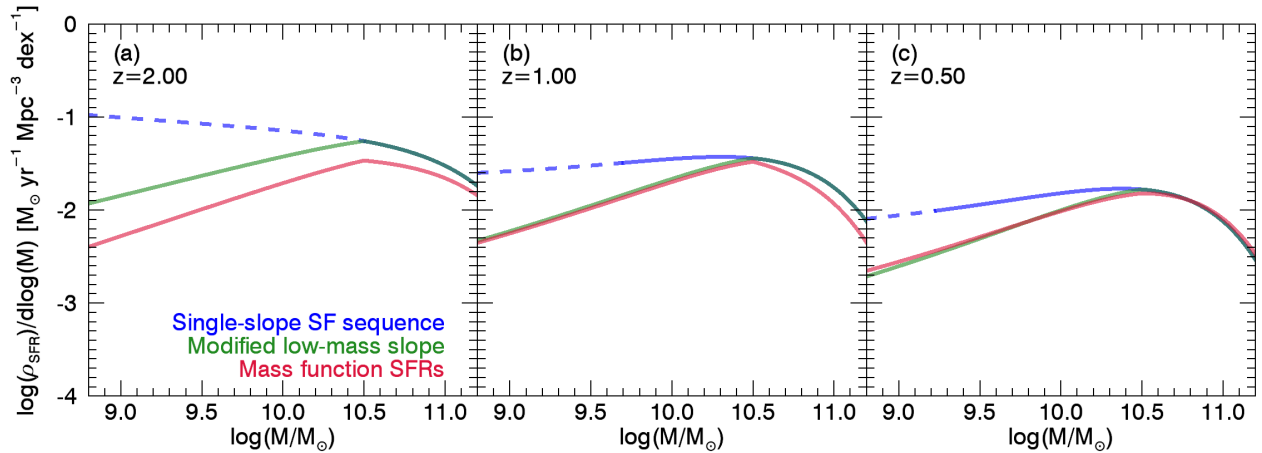}
\caption{The contribution to the global star formation rate density as a function of galaxy stellar mass. This is calculated by multiplying the abundance of star-forming galaxies in ZFOURGE by the star-forming main sequence.  Incompleteness in the observed star-forming sequence from {Whitaker} {et~al.} (2012) is indicated by the dashed line. Extrapolating the observed star-forming sequence results in a substantial contribution from low-mass galaxies to the star formation rate density, particularly at high redshift.}
\label{sfrd}
\end{center}
\end{figure*}

If the star formation rates are solely responsible for the discrepancy, even after fixing the low-mass slope to unity, star formation rates would have to decrease across all masses at $z>1$. The difference at $z=2$ can be read off of Figure \ref{mainplot}: star formation rates would have to decrease by $\sim 0.5$ dex at $\mathrm{log(\mathrm{M/M_{\odot}})}=9$, decreasing to $0.1$ dex at $\mathrm{log(\mathrm{M/M_{\odot}})}=11$ (again, before accounting for growth via mergers). The star-forming sequence reported in {Whitaker} {et~al.} (2012) represents the median star formation rate for star-forming galaxies. 

Notably, the median SFR is 0.1 dex below the average star formation rate for star forming galaxies ({Whitaker} {et~al.} 2014; {Kelson} 2014). Correcting for this offset does not change the results of this study-- indeed, it would only further increase the discrepancy between star formation rates and the growth of stellar mass. 

Some of the discrepancy at low masses can be alleviated by mergers decreasing the number density of galaxies. However, mergers also contribute to the stellar mass growth of galaxies at the high-mass end, which is implicitly reflected in the growth of the mass function. Thus, accounting for mass growth via mergers would increase the discrepancy between star formation rates and stellar mass growth at the high-mass end, and there will still exist a global difference between star formation rates and stellar mass growth.

We first ask whether this disagreement is particular to the star formation rates measured in the {Whitaker} {et~al.} (2012) star-forming sequence. We perform a comparison between the integrated star formation rate densities implied by star-forming sequences from the literature, shown in Figure \ref{litcomp} ({Oliver} {et~al.} 2010; {Kajisawa} {et~al.} 2010; {Peng} {et~al.} 2010; {Karim} {et~al.} 2011; {Whitaker} {et~al.} 2012). We use the abundance of star-forming galaxies from ZFOURGE to convert the star-forming sequence into star formation rate densities. The implied global star formation rates vary by a factor of two, but display similar redshift evolution. The similar redshift evolution between studies is due to the broad agreement in the redshift evolution of the normalization of the starforming sequence. The actual offsets are due to a combination of different normalizations and different slopes; crucially, changing the limits of integration has a significant effect on this comparison, due to the difference in observed slopes. These differences in slopes and normalization may come about due to different SFR indicators, different adopted conversions from luminosity to SFR, different dust corrections, selection effects, or different definitions of a star-forming galaxy ({Speagle} {et~al.} 2014). Since the redshift evolution differs little between studies, the relative rate of change may be well-constrained-- at least for galaxies with $\log(\mathrm{M/M_{\odot}})$ $>$ 10. Notably, the $\sim0.3$ dex study-to-study variation between star formation rates is comparable in size to the systematic decrease in star formation rates necessary to bring the evolution of stellar mass and star formation rates into agreement.

As many of the observable SFR indicators are primarily driven by radiation from massive stars, one way to decrease observed star formation rates is to postulate a top-heavy IMF (e.g., {van Dokkum} 2008). It has been claimed in the past that a top-heavy IMF is necessary to reproduce the properties of the submillimeter galaxy population ({Baugh} {et~al.} 2005; {Dav{\'e}} {et~al.} 2010), though recent studies may have resolved this tension ({Hayward} {et~al.} 2013). There is not enough systematic evidence for a top-heavy IMF to be compelling so far ({Bastian}, {Covey}, \& {Meyer} 2010).

Alternatively, systematic overestimation of stellar masses at high redshift could solve the tension with star formation rates. The cumulative redshift-dependent errors in the mass function necessary to be consistent with the modified star-forming sequence can be read directly from the bottom panels of Figure \ref{mainplot}: the error at high masses would have to be $0.2-0.3$ dex, and $0.4$ dex at low masses.  At the high-mass end, errors could potentially come from systematic errors in fitting the light profile ({Bernardi} {et~al.} 2013). At the low-mass end, stellar masses may be significantly overestimated due to emission line contributions to the observed flux. Fixing the stellar metallicity to the solar value is likely a poor approximation at high redshift, and may lower stellar mass estimates as well, particularly at the low-mass end ({Mitchell} {et~al.} 2013). It is also possible that the exponentially-decaying SFHs are poor fits to true SFHs of $z=2$ galaxies, which would decrease stellar mass at all masses ({Papovich} {et~al.} 2011).

In Figure \ref{litcomp}, we compare the integrated stellar mass density from the ZFOURGE survey to other results from the literature ({Marchesini} {et~al.} 2009; {Muzzin} {et~al.} 2013; {Moustakas} {et~al.} 2013). These studies agree to within a factor of two from $0<z<2.25$, but they show different evolution with redshift. In particular, for $1.5<z<2.25$, ZFOURGE shows less evolution with redshift than other surveys, which implies correspondingly lower star formation rates. If the integration limits are extended to lower stellar masses, the integrated stellar mass density in ZFOURGE shows even less evolution with redshift, due to relatively high abundance of low-mass galaxies at high redshift in ZFOURGE. 

This comparison shows that the uncertainties in the absolute value of stellar mass density in the literature are at least a factor of two, and this difference is redshift-dependent. These systematic redshift-dependent errors in stellar mass estimates may substantially contribute to the difference between the growth of stellar mass and the star formation rate at $z=2$.

We note that altering the IMF for low-mass stars as suggested by {Conroy} \& {van Dokkum} (2012) will affect star formation rates and stellar masses equally, and thus will not solve the tension between them.

\subsection{Which galaxies dominate the cosmic star formation rate density?}
Here, we examine the effects of different forms of the star-forming sequence on the global star formation rate. The mass-dependent contribution to the global star formation rate implied by the star-forming sequence, calculated via $n_{sf}\times SFR(M)$, is shown in Figure \ref{sfrd}.

\begin{figure}[t]
\begin{center}
\includegraphics[bb=100 30 570 700,scale=0.65]{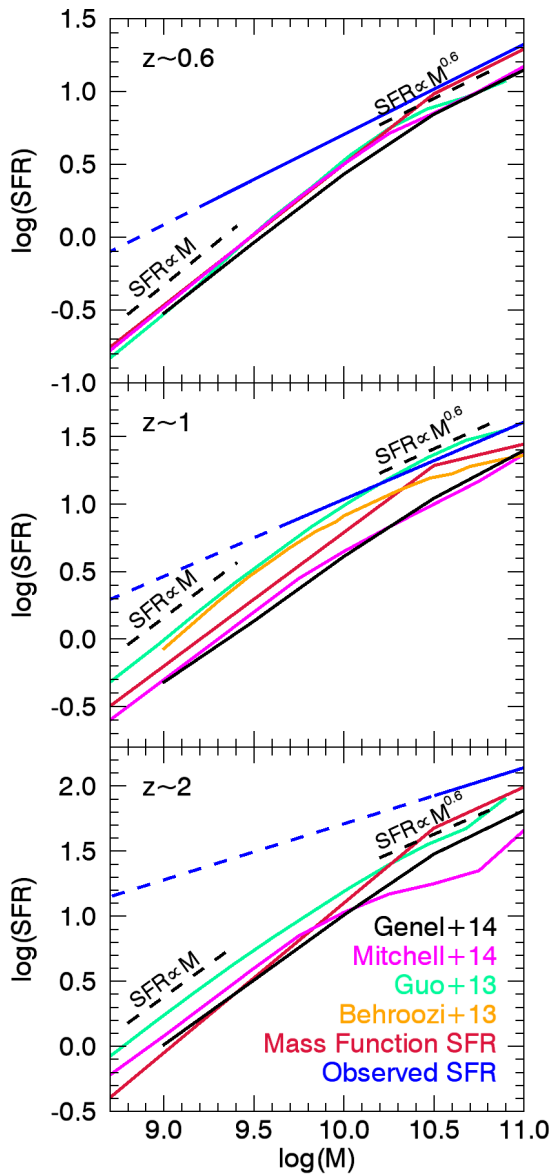}
\caption{The star-forming sequences from semi-analytical models ({Guo} {et~al.} 2013b; {Mitchell} {et~al.} 2014), abundance matching models ({Behroozi} \& {Silk} 2014, shown only in the middle panel), and hydrodynamical models ({Genel} {et~al.} 2014) also show a steep low-mass slope and a shallow high-mass slope. This may be a universal feature of models which reproduce the cosmic evolution of the stellar mass function.}
\label{theorycomp}
\end{center}
\end{figure}

With no modifications to the low-mass slope of the star-forming sequence, low mass galaxies dominate the cosmic star formation rate at $z=2$, and continue to contribute substantially to the cosmic star formation rate at lower redshifts. This is consistent with {Reddy} \& {Steidel} (2009), which suggests that sub-$L^*$ galaxies constitute up to 93\% of the unobscured UV cosmic star formation rate density at $2<z<3$. However, {Sobral} {et~al.} (2014) use data from a narrowband survey of H$\alpha$ emission at $z=2.23$ to argue that the contribution to the cosmic star formation density peaks at galaxies of $10^{10}$ M$_{\odot}$. This is in agreement with the $z\sim1$ star formation rate density measured from the ROLES survey ({Gilbank} {et~al.} 2010). Semi-analytical models predict a peak at a similar location, though at slightly higher stellar masses: $\sim$$10^{10.5}$ M$_{\odot}$, decreasing to 10$^{10}$ M$_{\odot}$ at $z=0$ ({Lagos} {et~al.} 2014).

\begin{figure*}[t!]
\begin{center}
\includegraphics[bb=200 120 380 700, scale=0.7,angle=90]{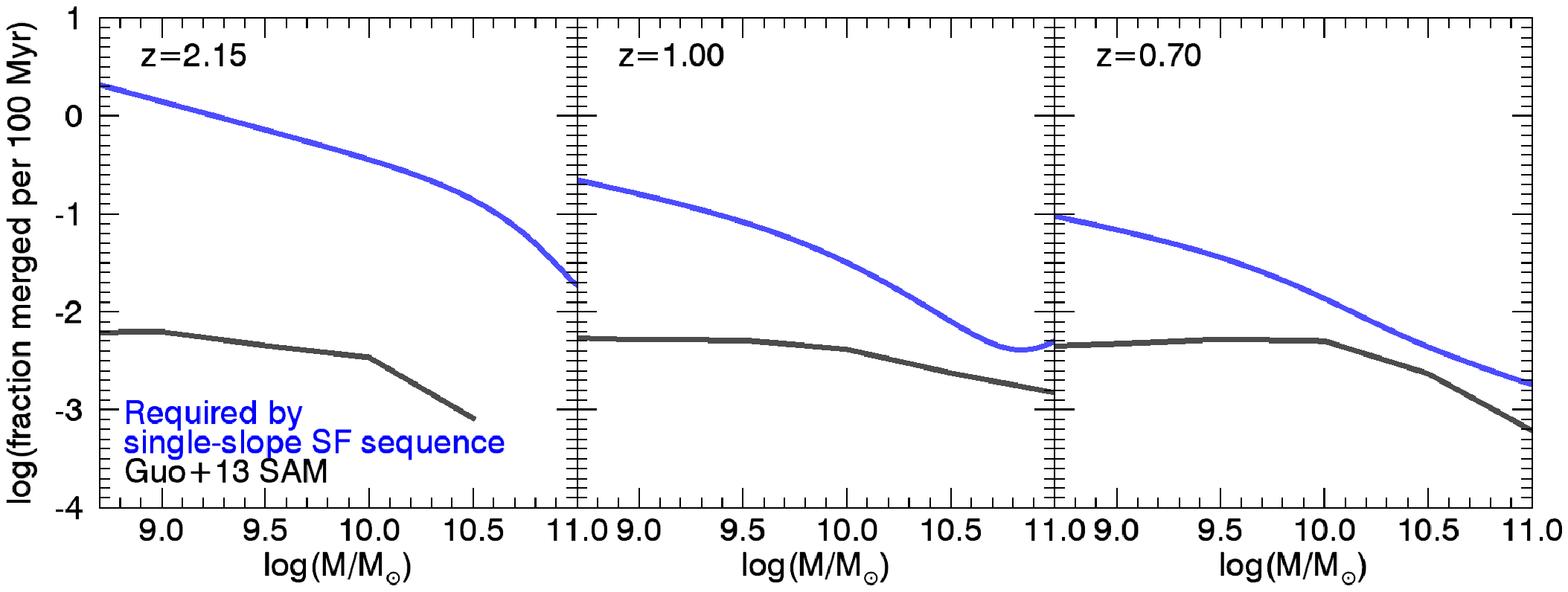}
\includegraphics[bb=180 120 460 700, scale=0.7,angle=90]{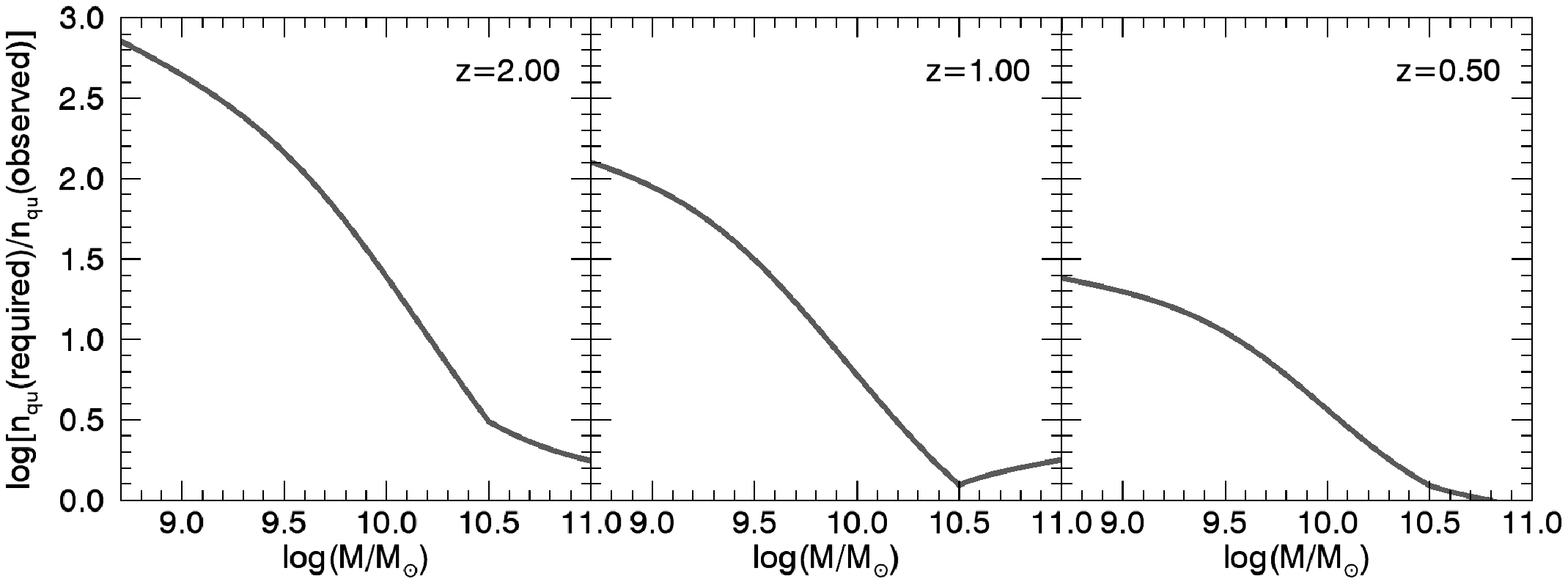}
\caption{Top panels: The merger rate required for a low-mass extrapolation of the star-forming sequence to be consistent with the evolution of the mass function. This is compared to the mass-dependent destruction rate of galaxies from the {Guo} {et~al.} (2013b) SAM. A mass-independent slope for the star-forming sequence implies merger rates that are orders of magnitude higher than predicted by dark matter models, particularly at $z=2$. Bottom panels: The factor by which the observed number density of quiescent galaxies must increase to make a low-mass extrapolation of the star-forming sequence consistent with the evolution of the mass function. At low masses and high redshifts, the number density of quiescent galaxies would have to be three orders of magnitude higher than what is observed. For reference, the quiescent mass function in ZFOURGE is complete to $\log(\mathrm{M/M_{\odot}})$ = $9$ at $z=2$.}
\label{sffrac_merge}
\end{center}
\end{figure*}

Adopting a steep low-mass slope, as suggested by our analysis, substantially decreases the implied contribution from low-mass galaxies, in agreement with {Gilbank} {et~al.} (2010) and {Sobral} {et~al.} (2014)-- though notably, the {Gilbank} {et~al.} (2010) star formation rate density falls off more sharply towards low stellar mass, due to differences between the ROLES and ZFOURGE stellar mass functions ({Gilbank} {et~al.} 2011). The peak of the cosmic star formation rate density depends on the exact location of the transition between high-mass slope and low-mass slope; in our model, this is fixed to $10^{10.5}$ M$_{\odot}$. The star formation rates implied by the evolution of the stellar mass function have an almost identical shape to the modified low-mass star-forming sequence, but with a $\sim0.3$ dex normalization offset at $z=2$. Both of the modified star-forming sequences presented in our study indicate that the contribution to the cosmic star formation peaks in galaxies of $10^{10}-10^{10.8}$ solar masses. An interesting parallel can be drawn with the star formation efficiency (defined as M$_*$/M$_{halo}$). Hydrodynamical simulations and abundance matching models indicate that the star formation efficiency peaks at a similar stellar mass $\log(\mathrm{M/M_{\odot}}) = 10.0-10.3$ for $0<z<4$ ({Behroozi}, {Wechsler}, \&  {Conroy} 2013b; {Genel} {et~al.} 2014).

\subsection{Comparison to galaxy formation models}
The observed star formation rates of galaxies at $z=1-3$ have long been in tension with theoretical expectations, which include semi-analytical, hydrodynamical, and semi-empirical models of galaxy formation ({Bouch{\'e}} {et~al.} 2010; {Firmani}, {Avila-Reese}, \&  {Rodr{\'{\i}}guez-Puebla} 2010; {Dav{\'e}}, {Oppenheimer}, \&  {Finlator} 2011; {Lilly} {et~al.} 2013; {Dekel} \& {Burkert} 2014; {Genel} {et~al.} 2014; {Mitchell} {et~al.} 2014). In models of galaxy formation, the specific star formation rate of galaxies roughly tracks the specific dark matter accretion rate ({Weinmann} {et~al.} 2012; {Lilly} {et~al.} 2013; {Mitchell} {et~al.} 2014; {Genel} {et~al.} 2014; {Peng} {et~al.} 2014), with a normalization offset of $\sim3$, required to match low-redshift observations ({Genel} {et~al.} 2014). However, between $z=1-3$, the observed star formation rates are higher than the model star formation rates by a factor of $\sim2$, despite matching the model star formation rates at both higher and lower redshifts. 

The discrepancy increases at low stellar masses, where galaxy formation models prefer a star-forming sequence with a slope of approximately unity, while some observations indicate a shallower slope ({Genel} {et~al.} 2014).  One solution is to de-couple galaxy formation and halo assembly at low stellar masses: {Weinmann} {et~al.} (2012) suggest this, arguing that observations indicate that the buildup of the number density of galaxies with $9.27<$ $\log(\mathrm{M/M_{\odot}})$$<9.77$ occurred much more recently than in the models. {Mitchell} {et~al.} (2014) advocate for a similar picture, arguing that the steep star-forming sequence in semi-analytical models must be adjusted at low stellar masses to match observations of a shallower slope.

In Figure \ref{theorycomp}, we compare the star-forming sequence in semi-analytical models ({Guo} {et~al.} 2013a; {Mitchell} {et~al.} 2014), stellar mass-halo mass abundance matching (SHAM) model ({Behroozi} \& {Silk} 2014), and hydrodynamical models ({Genel} {et~al.} 2014) to the star-forming sequence we derive from the ZFOURGE mass functions. They are strikingly similar in shape, with a slope of $\sim0.6$ at the high-mass end and a slope of $\sim1$ at the low-mass end. Considering that semi-analytical, hydrodynamical, and abundance-matching models must necessarily be consistent with the evolution of the stellar mass function, it is perhaps not surprising that they show similar behavior to the star formation rates derived from the observed redshift evolution of the stellar mass function.

We suggest that a mass-dependent slope for the star-forming sequence would ease the discrepancy between models and observations. A steep low-mass slope at $z>1$ is not ruled out by current studies: see the discussion in Section \ref{obs_sfseq}. Adopting a star-forming sequence with a slope close to unity at low masses means that low-mass galaxies grow in a self-similar fashion, implying that $\dot{M}/M=c(z)$, where $c(z)$ is a constant that depends only on redshift. Thus, the relative growth of low-mass galaxies would be independent of mass, but still dependent on redshift. The star formation rate of low-mass galaxies will then scale similarly to the specific dark matter accretion rate, which, for $\dot{M}\propto M^{\alpha}$, has $\alpha=1.1$ ({Neistein} \& {Dekel} 2008). 

\subsection{Alternatives to changing the low-mass slope: undetected quiescent galaxies and mergers}
\label{otherexp}
Enforcing a steep slope in the star-forming sequence at low stellar masses is one way to construct a consistent picture of the buildup of stellar mass. However, it is important to explore possible alternative explanations as well: namely, the merger rate and the possibility of a hidden population of quiescent galaxies. Higher merger rates will remove low-mass galaxies before they grow rapidly, while large populations of quiescent galaxies will decrease the average star formation rate as a function of mass.

We calculate the merger rate needed to produce a single-slope star-forming sequence. This is derived by growing the stellar mass function with the single-slope star-forming sequence, then computing the difference between the observed mass function and the mass function grown by star formation. This difference, divided by the time step, is the required rate of destruction by mergers. Note that while we neglect the galaxy growth due to mergers in this procedure, in practice, the growth rate due to mergers in our model is negligible for galaxies with stellar mass lower than 10$^{11}$ M$_{\odot}$.

The same effect can be caused by undetected quiescent galaxies. To calculate the required number density of undetected quiescent galaxies, we first calculate the growth of the stellar mass function in a single time step due to the single-slope star-forming sequence from {Whitaker} {et~al.} (2012). Then, we compare this growth rate to the growth rate in a single time step required to match the observed stellar mass function. The increase in the number of quiescent galaxies necessary to reconcile these two growth rates is inferred from Equation 6. 

Figure \ref{sffrac_merge} shows the merger rate necessary for the observed star-forming sequence to be consistent with the evolution of the mass function. We compare this directly to the merger rate from the {Guo} {et~al.} (2013b) semi-analytical model. At $z=2$, merger rates would have to be over two orders of magnitude higher than expected in order be consistent with  the observed star-forming sequence. This difference decreases with redshift, but even at $z=0.5$, it is necessary to increase the rate of low-mass mergers by an order of magnitude to match the growth of the mass function.
 
Figure \ref{sffrac_merge} also shows the factor by which the number density of quiescent galaxies would have to increase to bring star formation rates and the growth of stellar mass  into agreement. The number density of quiescent galaxies at $\log(\mathrm{M/M_{\odot}})=9$ would have to increase by almost three orders of magnitude at $z=2$. At $z<1$, it would need to increase by two orders of magnitude. This would indicate that galaxy surveys are missing the overwhelming majority of quiescent galaxies at $z=1$.

ZFOURGE is estimated to be mass-complete for quiescent galaxies down to $\log(\mathrm{M/M_{\odot}})$ = 9 at $z=2.25$. However, the ZFOURGE stellar mass-complete limit is estimated by comparing the magnitude of a single stellar population formed at $z=5$ to the magnitude limit of the survey. This calculation does not take into account the confounding effects of dust. If low-mass quiescent galaxies have a significant amount of dust at higher redshift, this would imply that the nominal mass-completeness of the survey is overestimated.

We conclude that while it is possible that a combination of merger rates and missing quiescent galaxies may be a factor in easing the tension between star formation rates and the growth of stellar mass, they are unlikely to be a dominant factor. The most likely solution remains a modification of the low-mass slope of the star-forming sequence.

\subsection{Model approximations}
\subsubsection{Scatter in star formation rates}
\label{scatter}
We model the growth of the stellar mass function by assuming all galaxies grow at the same rate. At fixed stellar mass, however, galaxies are a diverse population and display a variety of star formation rates ({van Dokkum} {et~al.} 2011). It is thus important to discuss whether the assumption of a single growth rate at fixed mass is a valid one.

It is possible to mathematically prove that scatter in star formation rates does not influence the evolution of the mass function, and that the only relevant parameter is the {\it average} star formation rate as a function of mass and time. Here, we sketch this proof briefly; the full proof will be presented in Franx et al. (in preparation). 

The evolution of the mass function due to star formation is given by the following equation:
\begin{equation}
\label{galaxycont}
 \frac{\partial \psi(M,t)}{\partial t} = - \frac{\partial}{\partial M}\left[ \psi(M,t) \dot{M} \right]
\end{equation}
with $\dot{M}$ defined as the average star formation rate at mass $M$. This continuity equation follows immediately from the conservation of galaxies, and was first presented by {Drory} \& {Alvarez} (2008). The equation is identical to the continuity equation in stellar dynamics (e.g. {Binney} \& {Tremaine} (2008), equation 4.204). In this case, the mass $M$ is the x-position, and the star formation rate is the velocity.

As further discussed by Franx et al (in preparation), the consequence is that the evolution of the mass function due to star formation at any $M$ is determined by  $\dot{M}$ only, and is the same for all models with a given $\psi(M)$ and $\dot{M}$. Hence the scatter in the star formation at a given $M$ does not influence the evolution of the mass function - as long as  $\dot{M}$ is given. It is not important on what timescale galaxies may move below or above the average. As shown by Franx et al (in preparation), another way to see this quickly is by assuming that scatter is caused by two populations of  galaxies with two different star formation rates. In Appendix B, we show that sum of the two populations will satisfy the continuity equation, just like the two populations separately. This argument can be extended to an arbitrary number of subpopulations.

Thus, any galaxy population for which the mean star formation rate as a function of mass is identical will evolve in the same fashion. The actual distribution of star formation rates does not affect the evolution of the stellar mass function: as long as the mean star formation rate as a function of mass is identical in any given distribution at {\it every time step}, the mass function will evolve in an identical fashion.

%
%

\subsubsection{Modeling mass loss from galaxies}
\label{mloss}
We adopt an instantaneous stellar mass loss model. In reality, the mass loss rate of galaxies is time-dependent, with the majority of it occurring in the first hundred million years. The efficiency of the instantaneous approximation will thus scale inversely with the specific star formation rate of galaxies, meaning that low-mass galaxies at high redshift are most affected by this approximation.

Stellar mass functions are tabulated using the current mass in stars and stellar remnants, thus implicitly taking this effect into account. To account for this in our model, the fraction of mass returned to the ISM, $R$, would have to be modeled as SFH-dependent. This would likely elevate the implied star formation rates at low masses by $\sim 20\%$. This is not a dominant effect, but would contribute towards reconciling stellar mass growth and star formation rates at high redshift and low masses.

\section{Conclusion}
In this study, we have examined the connection between the observed star-forming sequence and the observed redshift evolution of the stellar mass function. We have constructed a smooth parameterization of the growth of the stellar mass function from ZFOURGE and SDSS-GALEX data, and compared this growth to the growth implied by the observed star-forming sequence. We find that a simple extrapolation of the observed slope to low stellar masses is inconsistent with the observed evolution of the mass function. We conclude from this comparison that one or all of the following must be true: (1) the star-forming sequence steepens at low masses, (2) the destruction rate of low-mass galaxies by mergers is very high, (3) there is a dominant population of low-mass quiescent galaxies missing from high-redshift surveys. We use merger rates from semi-analytical models to show that the merger rate of low-mass galaxies is several orders of magnitude too low to solve the issue. We also show that there would have to be several orders of magnitude more quiescent galaxies than observed at $z=2$ to solve the issue, which is unlikely to be true in the mass-complete regime of stellar mass surveys. We thus conclude that, in order to decrease the star formation rates for galaxies with mass $\sim10^9$ M$_{\odot}$, the star-forming sequence must steepen at low masses. This conclusion is supported by the observations of {Whitaker} {et~al.} (2014).

We show that a mass-dependent slope for the star-forming sequence makes the global star formation rate more consistent with the growth of stellar mass density. However, even with a steeper star-forming sequence, there is a discrepancy of $\sim0.3$ dex between the growth of stellar mass density and the global star formation rates at high redshift. Comparing different studies in the literature, we find that there is an inter-study difference of $\sim0.3$ dex in both the implied global star formation rates and the growth of the stellar mass density, and the redshift evolution of these quantities can differ substantially between studies. It is thus unclear whether measurements of star formation rates, stellar masses, or both need to change in order to tell a consistent story.

Future efforts to firmly establish the observational uncertainties in star formation rates and stellar masses will be crucial to resolving this discrepancy.

\acknowledgements
We thank the anonymous referee for their insightful suggestions and attention to details, which have greatly improved the quality of the paper. Support from NASA Grant NNX11AB08G is gratefully acknowledged.

\appendix
\section{Mass-Dependent Merger Rates from Semi-Analytical Models}
Our fiducial model, presented in the main body of the paper, uses the destruction rate of galaxies from the {Guo} {et~al.} (2013b) SAM, and assumes that the resulting mass growth occurs in 1:10 ratio mergers. This approximation has the advantage of enforcing mass conservation within our model.

However, this approximation also introduces several issues. Firstly, by simplifying all mergers into 1:10 mass ratio mergers, the mass growth rate of a galaxy of mass M is determined only by the destruction rate at 0.1*M, modulated by the ratio of number densities. In particular, this underestimates the mass growth due to mergers in high mass galaxies, which don't just accrete intermediate-mass galaxies but also accrete a significant fraction of low-mass galaxies. Secondly, galaxies below the mass cutoff of our simulation (10$^6$ M$_{\odot}$) also deposit stellar mass into more massive galaxies, but this is not reflected in our prescriptions.

\begin{figure}[]
\begin{center}
\includegraphics[bb=120 30 410 750, scale=0.5]{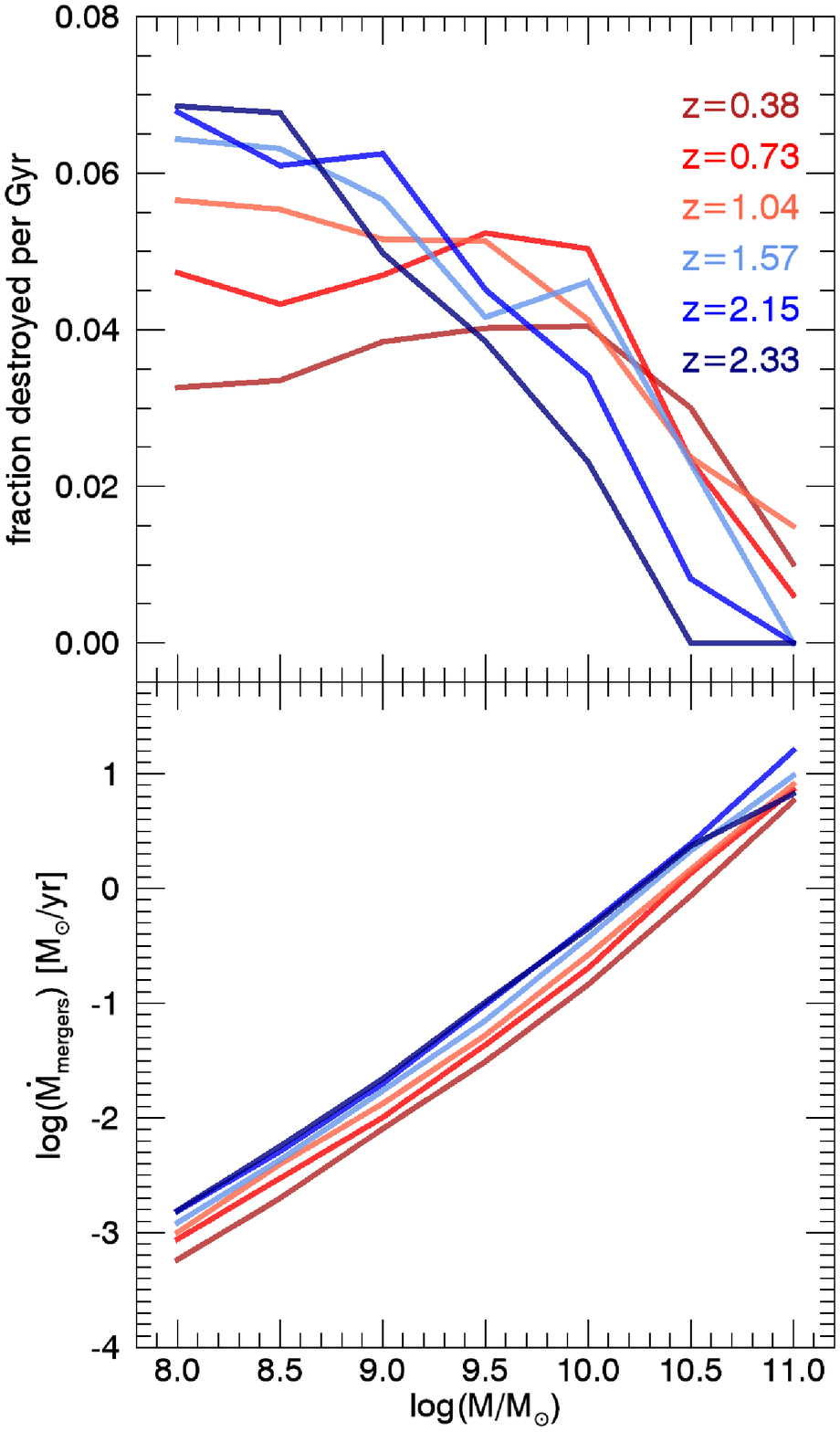}
\caption{Mass-dependent destruction rates and growth rates, measured directly from the {Guo} {et~al.} (2013b) SAM as a function of redshift. At $z\sim2$, low-mass galaxies are being rapidly destroyed, but this rate of destruction decreases as redshift decreases. The opposite trend is seen at higher masses, whereby the rate of destruction for higher-mass galaxies increases with decreasing redshift. The growth rate due to mergers is a smooth and continuous function of mass.}
\label{samdirect}
\end{center}
\end{figure}

To address these issues, we directly measure the mass growth of galaxies due to mergers in the {Guo} {et~al.} (2013b) SAM, and apply them to the stellar mass function. The destruction rates and growth rates as a function of mass are shown in Figure \ref{samdirect}. The growth due to mergers is a smooth function of mass, and decreases smoothly and slowly as redshift decreases. The destruction rate due to mergers increases with redshift at the low-mass end and decreases with redshift at the high-mass end, such that a destruction rate which is relatively flat with mass at $z=0.38$ becomes a destruction rate which is quite steep with mass at $z=2.33$. We note that the Millenium-II simulation probes a relatively small volume, such that at the highest masses, there are only a few galaxies from which to measure merger growth rates. The accuracy of the growth and destruction rates at the highest masses are thus strongly limited by both Poisson noise and cosmic variance.

\begin{figure*}
\begin{center}
\includegraphics[bb=90 100 470 725, scale=0.7,angle=90]{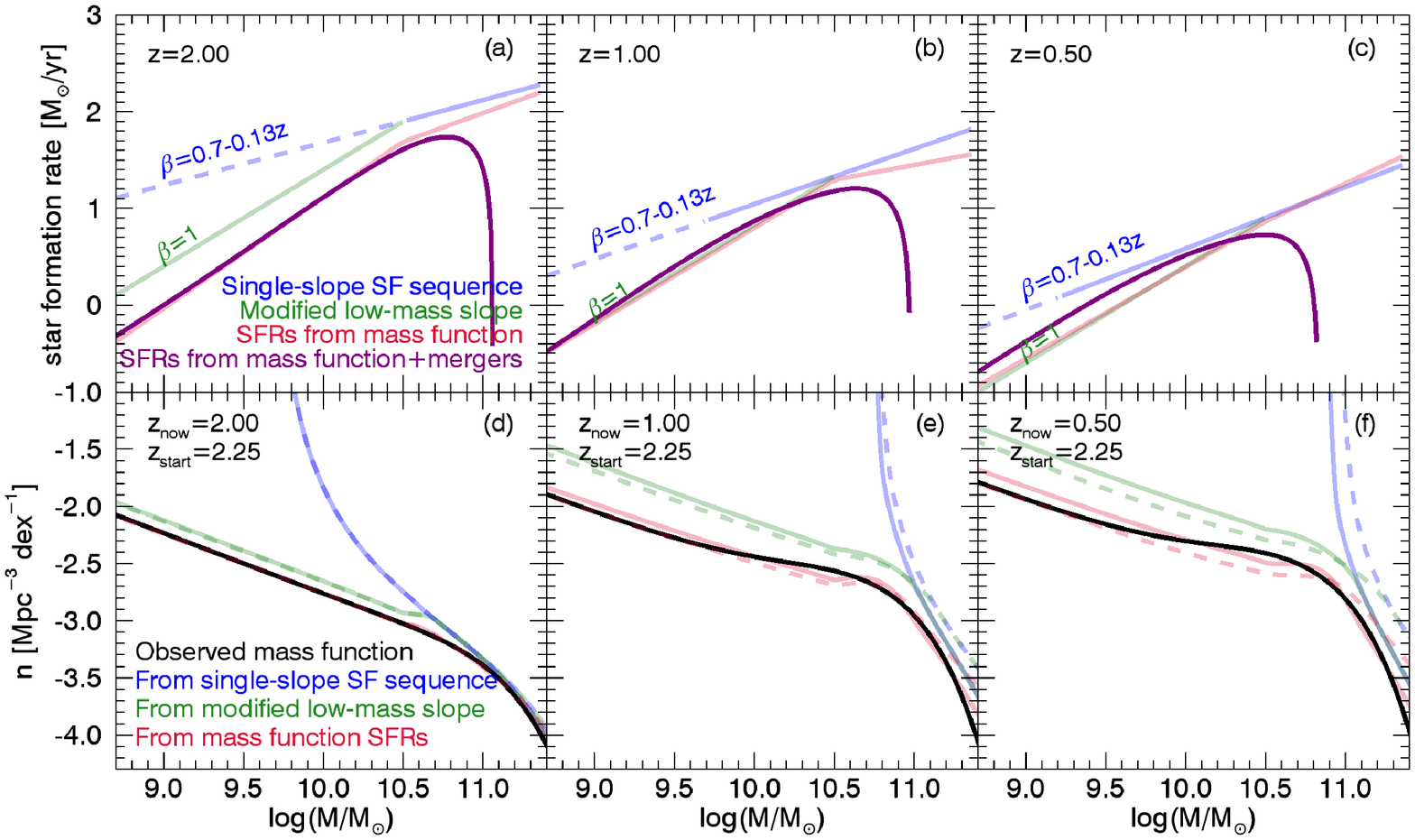}
\caption{Identical to Figure 3 in the main text, except the dashed curves in the lower three panels now represent mass functions modulated by the mass-dependent destruction rates and growth rates from the {Guo} {et~al.} (2013b) SAM. These can be compared directly to the solid curves, which do not include the effects of mergers. The effects of mergers in these calculations are relatively small at low masses. The purple line represents the star-forming sequence required to make the evolution of the stellar mass function, modulated by the SAM merger rates, match the observed evolution of the stellar mass function. The three models presented previously in the main body of the paper are here shown with faded colors.}
\label{mainplotappend}
\end{center}
\end{figure*}

We interpolate between the measurements in Figure \ref{samdirect} and apply both directly to the growth of the mass function. The resulting growth of the mass function, when combined with the three parameterizations of the star-forming sequence described in Section \ref{dif_sfseq}, is shown by the dashed lines in Figure \ref{mainplotappend}. The main difference as compared to our preferred model is the growth of galaxies with $\log(\mathrm{M/M_{\odot}})$ $>$ 11, which increases substantially when using the SAM growth rates. Lower-mass galaxies experience less growth due to mergers, but as their growth was already dominated by star formation, this has little effect on the evolution of the mass function.

\begin{figure}[]
\begin{center}
\includegraphics[bb=160 70 410 750, scale=0.5]{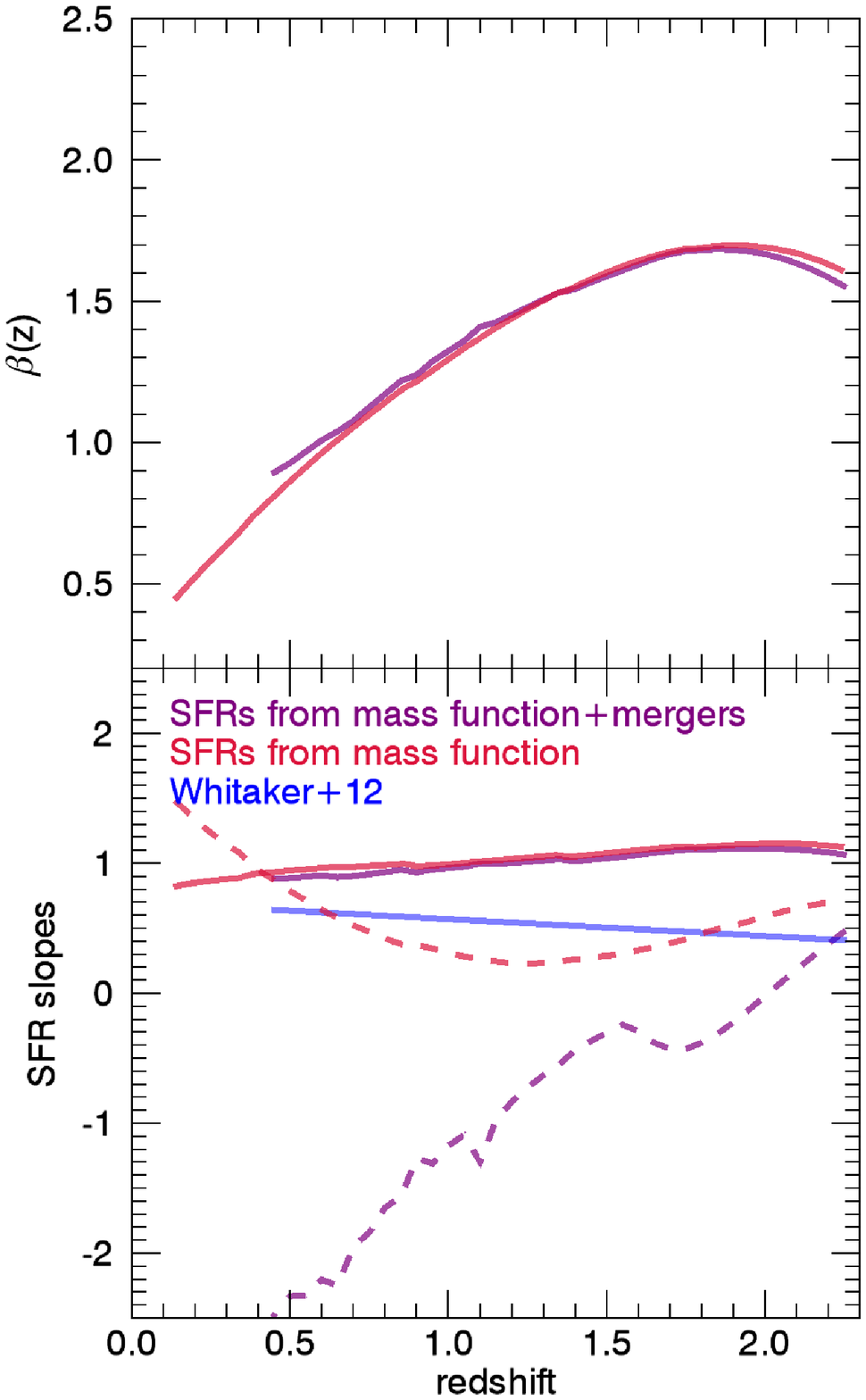}
\caption{The upper panel shows the redshift evolution of the normalization for the double-power law fits to the star-forming sequence, while the lower panel shows the same for the low-mass slopes (solid line) and high-mass slopes (dashed line). The red line is the star-forming sequence derived from the mass function as described in Section \ref{zfourgesfr}, while the purple line is done in the same way, but after applying the SAM-measured merger rates. The slope of the star-forming sequence becomes an even stronger function of mass after including mergers.}
\label{bpl_append}
\end{center}
\end{figure}

The main result of this study is that the growth of the mass function requires a mass-dependent slope in the star-forming sequence. This is demonstrated in Section \ref{zfourgesfr}, where we derive a star-forming sequence which is consistent with the evolution of the stellar mass function, and show that it has two distinct slopes. We now repeat this process of deriving a consistent star-forming sequence, but now perform this fit ${\it after}$ including the effects of mergers. This star-forming sequence is shown in purple in Figure \ref{mainplotappend}. Note that it is not a functional fit that is shown (as is true for the star formation rates derived from the mass function without mergers), but the actual inferred star formation rates.

Including the mergers when fitting for the star-forming sequence slightly increases the star formation rates at low masses, while decreasing them substantially at high masses. This is due to the fact that the main effect of mergers at low mass is the destruction of galaxies, while the main effect of mergers at high mass is additional mass growth. In fact, applying the {Guo} {et~al.} (2013b) SAM merger growth rates to the high-mass end is more than sufficient to model the growth of high-mass galaxies in the ZFOURGE mass function: presumably, the merger growth rates in the real Universe are lower than those measured from the SAM.

We also fit the merger-modulated star formation rates with a double power law with the same form as Equation \ref{brokenpowerlaw}. The redshift evolution of the normalization and the slopes is shown in Figure \ref{bpl_append}. Crucially, the evidence for a mass-dependent slope is not altered by including the growth rate due to mergers. In fact, this conclusion is strengthened, though this highlights an important point: deriving the high-mass slope of the star-forming sequence is very sensitive to the merger rate (conversely, the low-mass slope is very stable to the merger rate).

However, using merger growth rates measured directly from the SAM violates mass conservation. Specifically, the {Guo} {et~al.} (2013b) mass functions have a higher number density of low-mass galaxies than is observed in ZFOURGE, which means that for the same rate of destruction of galaxies, the merger growth rates from the semi-analytical model will be higher than the simple 1:10 merger growth rates assumed in our model. This results in mass growth at the high-mass end which is more than enough to explain the growth of the mass function. This extra growth is reflected in the high-mass drop off in the merger-modulated star formation rates from the stellar mass functions, visible in Figure \ref{mainplotappend}.

\section{Linearity of the Continuity Equation}
In this section, we prove that Equation \ref{galaxycont} presented in the text is linear, and thus can be generalized to the entire galaxy population.

First, assume that we have two population of galaxies, each with a different dependence of star formation rate with stellar mass. The continuity equation for each individual population is
$$
\frac{\partial \psi_i(M,t)}{\partial t} = - \frac{\partial}{\partial M}\left[ \psi_i(M,t) \langle \dot{M}_i \rangle \right]
$$
where the subscript i is either 1 or 2 for the two populations.

We sum both sides of the continuity equation, as defined for galaxy populations 1 and 2. For the sum of the left hand terms, we obtain:
$$
\frac{\partial \psi_1(M,t)}{\partial t} + \frac{\partial \psi_2(M,t)}{\partial t} = \frac{\partial [\psi_1(M,t)+\psi_2(M,t)]}{\partial t}
$$
and for sum of the right hand terms, we obtain:
$$
 - \frac{\partial}{\partial M}\left[ \psi_1(M,t)\langle \dot{M}_1 \rangle \right] - \frac{\partial}{\partial M}\left[ \psi_2(M,t) \langle \dot{M}_2 \rangle \right] =  
 $$
 $$
 - \frac{\partial}{\partial M} \left[\psi_1(M,t) \langle \dot{M}_1 \rangle + \psi_2(M,t) \langle \dot{M}_2 \rangle\right] 
 $$
 Now, define $\langle \dot{M}_i \rangle$ to be the average star formation rate of the combined galaxy population, and $\psi_{tot}(M,t)$ to be the total mass function:
 $$ 
 \psi_{tot}(M,t) = \psi_1(M,t)+\psi_2(M,t)
 $$
 $$
\langle \dot{M}_{tot} \rangle = \frac{\psi_1(M,t)\langle \dot{M}_1 \rangle+ \psi_2(M,t)\langle \dot{M}_2 \rangle}{\psi_{tot}(M,t)}
 $$
 Substitute these into the left and right hand sums, and we recover the original equation:
$$
\frac{\partial \psi_{tot}(M,t)}{\partial t} = - \frac{\partial}{\partial M}\left[ \psi_{tot}(M,t) \langle \dot{M}_{tot} \rangle \right]
$$
Thus, the continuity equation for the evolution of the galaxy population is linear.


\end{document}